\documentclass[
]{jss}

\usepackage{orcidlink,thumbpdf,lmodern}

\usepackage[utf8]{inputenc}

\author{
Adam Rauh\\University of Michigan \And In Song Kim\\MIT \And Kosuke
Imai\\Harvard University
}
\title{\pkg{PanelMatch}: Matching Methods for Causal Inference with
Time-Series Cross-Sectional Data}

\Plainauthor{Adam Rauh, In Song Kim, Kosuke Imai}
\Plaintitle{PanelMatch: Matching Methods for Causal Inference with
Time-Series Cross-Section Data}
\Shorttitle{\pkg{PanelMatch}}

\Abstract{
Analyzing time-series cross-sectional (also known as longitudinal or
panel) data is an important process across a number of fields, including
the social sciences, economics, finance, and medicine.
\textbf{PanelMatch} is an R package that implements a set of tools
enabling researchers to apply matching methods for causal inference with
time-series cross-sectional data. Relative to other commonly used
methods for longitudinal analyses, like regression with fixed effects,
the matching-based approach implemented in \textbf{PanelMatch} makes
fewer parametric assumptions and offers more diagnostics. In this paper,
we discuss the \textbf{PanelMatch} package, showing users a recommended
pipeline for doing causal inference analysis with it and highlighting
useful diagnostic and visualization tools.
}

\Plainkeywords{matching, observational studies, panel data, causal
inference, weighting}

\Address{
    Adam Rauh\\
    University of Michigan\\
    Department of Political Science\\
505 South State Street, Ann Arbor, MI 48109\\
  E-mail: \email{amrauh@umich.edu}\\
  
      In Song Kim\\
    MIT\\
    Department of Political Science\\
77 Massachusetts Avenue, Room E53-407, Cambridge, MA, 02142\\
  E-mail: \email{insong@mit.edu}\\
  
      Kosuke Imai\\
    Harvard University\\
    Department of Government and Department of Statistics\\
1737 Cambridge Street, Cambridge, MA, 02142\\
  E-mail: \email{imai@harvard.edu}\\
  
  }

\usepackage{amsmath} \usepackage{graphicx} \usepackage{booktabs} \graphicspath{ {./figs/} } \usepackage{amssymb,amsmath, bm} \usepackage{fancyvrb} \usepackage{float} \usepackage{caption} \usepackage[large]{subfigure} \usepackage{tikz} \usepackage{inputenc}

\usetikzlibrary{shapes,arrows,trees,fit,positioning,shapes.misc, calc}
      \newcommand\spacingset[1]{\renewcommand{\baselinestretch}{#1}\small\normalsize}

\begin{document}

\section{Introduction}\label{introduction}

Estimating causal effects from time-series cross-sectional data is
crucial in various fields including political science, sociology,
economics, finance, and medicine. For instance, scholars might be
interested in exploring how changes like increasing the minimum wage
affect employment \citep{CardKrueger1994}, how transitioning to
democracy impacts GDP \citep{acemoglu_democracy_2017}, or how
introducing greenspaces influences public health
\citep{branas2011difference}. In these studies, treatments are applied
at different times across various units, and the outcomes are also
measured at different times.

The estimation of causal quantities from panel data is commonly
conducted using linear regression models with fixed effects. In
\textsf{R}, the \textbf{stats} package offers tools for basic analyses,
complemented by several additional packages that enhance these
capabilities \citep{stats-package}. The \textbf{plm} package is renowned
for its comprehensive functions that aid in data cleaning, fitting
linear models, and testing model specifications, specifically tailored
for panel data \citep{croissant2008panel}. The \textbf{lfe} and
\textbf{fixest} packages also provide highly efficient implementations
of fixed effects models and other model types, featuring user-friendly
interfaces \citep{RJ-2013-031, fixest}. Although this approach is widely
used to account for time-invariant unit-specific and/or unit-invariant
time-specific unobserved confounders, it is important to note that it
relies on parametric assumptions, which may not always be appropriate
for modeling the counterfactual outcomes in various settings
\citep{imai2021use, imaikim19}.

Much progress has been made recently to address these issues. For
instance, scholars have developed various methods based on the
difference-in-differences (DiD) identification strategy for estimating
causal quantities of interest: the \textbf{did2s} package, based on the
method by \citet{gardner2022two}, handles group-based heterogeneous
treatment effects and provides efficient, intuitive functions for
estimation \citep{RJ-2022-048}. The \textbf{didimputation} and
\textbf{staggered} packages, implementing methods by
\citet{borusyak2024revisiting} and \citet{roth2023efficient},
respectively, adapt two-way fixed effects estimators for staggered
treatment contexts \citep{didimputation, RothSantAnna2021}. The
\textbf{did} package, using the method by \citet{did-method}, allows for
causal effect estimation in settings with more than two time periods
\citep{did-package}. Additionally, \textbf{DRDID} enables users to
obtain doubly-robust treatment effect estimates \citep{drdid}.

In this article, we introduce and discuss
\textbf{PanelMatch}.\footnote{It is available for download from the Comprehensive \textsf{R} Archive Network (CRAN) at \url{https://CRAN.R-project.org/package=PanelMatch}}
\textbf{PanelMatch} contributes to the growing set of software packages
and methods available for causal inference with longitudinal data.
Specifically, \textbf{PanelMatch}, building on the approach of
\citet{imai2021matching}, enables the application of matching methods to
causal inference with panel data that have binary treatments. Matching
is a well-regarded nonparametric covariate adjustment methodology that
reduces model dependence in causal inference and is both intuitive and
accessible \citep{ho2007matching, stuart2010matching}. Unfortunately,
traditional matching methods are limited to cross-sectional data.

Trajectory balancing and synthetic control methods, such as those in the
\textbf{tjbal}, \textbf{Synth}, \textbf{gsynth}, and \textbf{synthdid}
packages, provide similar functionalities. However, they often come with
additional structural assumptions
\citep{hazlett2018trajectory, Synth, gsynth, arkhangelsky2021syntheticdifferencedifferences}.
For instance, many synthetic control methods require exactly one treated
unit, a large number of control units, or fixed treatment statuses
\citep{abadie2010synthetic, doudchenko2016balancing, ben2019synthetic, xu2017generalized}.

In contrast, \textbf{PanelMatch} allows for multiple units to be treated
at any point in time, and these units can change their treatment status
multiple times over the course of the study. Additionally, the proposed
methodology is effective even with a relatively small number of time
periods. The package offers a variety of diagnostic and visualization
tools that significantly extend what \citet{imai2021matching} initially
proposed. These include the ability to specify various causal quantities
of interest, visualize weights across different refinement methods,
perform placebo tests, and estimate treatment effects at the individual
matched set level.

\textbf{PanelMatch} is less dependent on model specifications, making it
more robust against model misspecification compared to regression-based
approaches. It is suitable for a wide range of contexts involving panel
data, without imposing any restrictions on the staggering or reversion
of treatment or the number of treated and control units---limitations
often present in existing packages. However, \textbf{PanelMatch} can
only be applied to contexts with binary treatment data, as is the case
with many existing methods and packages. Finally, \textbf{PanelMatch}
offers useful diagnostic and visualization tools that help validate the
credibility of results. Therefore, \textbf{PanelMatch} also enhances the
broader array of software tools available for managing and analyzing
panel data, including \textbf{panelr} and \textbf{PanelView}
\citep{panelr, mou2022panelview}.

\subsection{Package Overview}\label{package-overview}

\textbf{PanelMatch} is designed so that users follow a three step
process with corresponding functions. Table \ref{tab:pipeline}
summarizes this process, along with guiding considerations. First, users
identify matched sets of treated and control observations. Users begin
this step by preparing their data with \texttt{PanelData()} and then
visualizing the distribution of treatment across units and time using
the \texttt{DisplayTreatment()} function. Visualization will help
researchers examine the patterns of treatment variation in their data.
Users should then choose a quantity of interest for their desired
application. Given this selection, users will create sets of matched
treated and control observations with the \texttt{PanelMatch()}
function.

Second, users refine their matched sets by calculating weights for
control units in each matched set to obtain better covariate balance
between treated and matched control units. Many options are available to
fine-tune this process: the package contains implementations for various
matching and weighting methods based on propensity scores and
Mahalanobis distance. Users then calculate and visualize covariate
balance with a provided set of functions to evaluate the quality of
their matching and refinement.

Finally, if these diagnostics produce satisfactory covariate balance,
users compute estimates and standard errors using the
\texttt{PanelEstimate()} function. The validity and interpretation of
these estimates can be assessed by utilizing a final set of diagnostic
functions.

\begin{table}[]
\resizebox{\textwidth}{!}{
\begin{tabular}{@{}lll@{}}
\toprule
\textbf{Step}                                & \textbf{Package Function(s)}                                                                                          & \textbf{Suggested Considerations}                                                                                                                                                                                                                                    \\ \midrule
\large \textbf{1.} \begin{tabular}[c]{@{}l@{}} \textbf{Identify matched sets of} \\\textbf{treated and control observations} \end{tabular} &                                                                                            &                                                                                                                                            \\ \midrule
1.1 Prepare data & \texttt{PanelData()}                                                                                           & \begin{tabular}[c]{@{}l@{}} \textbullet\ Set unit ID, time ID, treatment, and \\ \hspace{1.1em}outcome variables \\ \textbullet\ Check for errors related to \\ \hspace{1.1em}data types and structure\end{tabular}                                                                                                                                                 \\ \midrule
1.2 Visualize treatment distribution & \texttt{DisplayTreatment()}                                                                                           & \begin{tabular}[c]{@{}l@{}} \textbullet\ Check variation across units, time\\  \textbullet\ If variation inadequate, stop\end{tabular}                                                                                                                                                 \\ \midrule
1.3 Determine quantity of interest   &                                                                                                            & \begin{tabular}[c]{@{}l@{}} \textbullet\ Use substantive knowledge\\ \textbullet\ ATT and ART are often appropriate\end{tabular}                                                                                                                                                     \\ \midrule
1.4 Create matched sets          & \texttt{PanelMatch()}                                                                                                 & \begin{tabular}[c]{@{}l@{}} \textbullet\ Determine appropriate lag window, lead window\end{tabular}                                                                                                                                                              \\ \midrule
1.5 Evaluate results and diagnostics          & \begin{tabular}[c]{@{}l@{}}\code{plot()}, \code{summary()}\end{tabular} & \begin{tabular}[c]{@{}l@{}} \textbullet\ Check size(s), number of matched sets\\ \textbullet\ If matched sets too small or \\\hspace{1.1em}too few, return to Steps 1.1-1.4\end{tabular}                                                                                                                   \\ \midrule
\large \textbf{2.} \textbf{Refine matched sets} &                                                                                            &                                                                                                                                            \\ \midrule
2.1 Apply refinement method to matched sets &               \texttt{PanelMatch()}                                                                             &  \begin{tabular}[c]{@{}l@{}} \textbullet\ Determine appropriate refinement method and\\ \hspace{1.1em}variables for refinement calculations \end{tabular}                                                                                                                                           \\ \midrule
2.2 Evaluate results and diagnostics & \begin{tabular}[c]{@{}l@{}}\code{get\_covariate\_balance()},\\ \code{plot()}, \code{summary()} \end{tabular}                                                                                           &     \begin{tabular}[c]{@{}l@{}} \textbullet\ Check covariate balance, \\\hspace{1.1em}which should be below .2 SDs\\ \textbullet\ If covariate balance too high,\\\hspace{1.1em}return to Step 2.1\end{tabular}                                                                                                                                       \\ \midrule

\large \textbf{3.} \textbf{Calculate estimates and standard errors} &                                                                                            &                                                                                                                                            \\ \midrule
3.1 Estimate quantity of interest    & \texttt{PanelEstimate()}                                                                                              & \begin{tabular}[c]{@{}l@{}} \textbullet\ Consider appropriate method of\\ \hspace{1.1em}calculating standard errors\end{tabular}                                                                                                                                                        \\ \midrule
3.2 Evaluate results and diagnostics          & \begin{tabular}[c]{@{}l@{}}\code{plot()}, \code{summary()},\\ \code{get\_set\_treatment\_effects()},\\ \code{placebo\_test()}\end{tabular}       & \begin{tabular}[c]{@{}l@{}} \textbullet\ Check distribution of set-level \\ \hspace{1.1em}effects, standard error size,\\\hspace{1.1em}placebo test results (if applicable)\\ \textbullet\ If distribution non-normal, \\\hspace{1.1em}standard errors too large,\\ \hspace{1.1em}or placebo test fails, return to previous steps\end{tabular} \\ \bottomrule
\end{tabular}}
\caption{\label{tab:pipeline} \textbf{Suggested Steps, Associated Functions, and Considerations for Using PanelMatch}.}
\end{table}

\subsection{Running Example}\label{running-example}

Example code throughout this paper will use a data set originally
analyzed by \citet{acemoglu_democracy_2017} who examined the question of
whether or not democratization causes economic growth. The
\textbf{PanelMatch} package includes a modified and simplified version
of this data set, which will be adequate for running all of the code
demonstrated in this article. This data set -- available as a
\texttt{data.frame} object named \texttt{dem} -- contains country-year
observations. See Table \ref{tab:acemoglu-toy} for a complete
description of the variables contained in this data set.

\begin{table}[]
\resizebox{\textwidth}{!}{
\centering
\begin{tabular}{@{}lll@{}}
\toprule
Variable Name & Description                                                                                                                                                                    & Data Type        \\ \midrule
\texttt{wbcode2}       & World Bank Country Code: identifies units, which are countries                                                                                                                 & Integer          \\
\texttt{year}          & Time variable                                                                                                                                                                  & Integer          \\
\texttt{dem}           & \begin{tabular}[c]{@{}l@{}}Treatment variable: describes status of a country in a given year as\\ democracy (1) or autocracy (0). Coded by Acemoglu et al. (2019)\end{tabular} & Integer \\
\texttt{y}             & Logged real GDP per capita                                                                                                                                                     & Numeric          \\
\texttt{tradewb}       & Trade volume as a fraction of GDP                                                                                                                                              & Numeric          \\ \bottomrule
\end{tabular}}
\caption{\label{tab:acemoglu-toy} \textbf{Names and Descriptions of Variables in Included Data}. The package includes a subset of the data used by Acemoglu et al. (2019). The names and descriptions of variables found in this data set are shown in the table. All example code shown in this article can be run using the included data.}
\end{table}

\section{Identifying Matched Sets}\label{identifying-matched-sets}

The first step of our analysis is the identification of matched sets of
treated and control observations. Users should begin by preparing their
data with the \texttt{PanelData()} function. \texttt{PanelData()}
conducts a number of error checks on the data, balances the panel, and
creates a \texttt{PanelData} object which stores the time identifier,
unit identifier, treatment, and outcome variables. Storing this metadata
simplifies the interface at later stages, so users do not need to
repeatedly specify these important variables.

The package provides built in \texttt{plot()}, \texttt{print()}, and
\texttt{summary()} methods for examining \texttt{PanelData} objects. The
code chunk below creates a \texttt{PanelData} object, and then leverages
the \texttt{print()} and \texttt{summary()} methods to inspect the
results. The \texttt{print()} method enables users to easily examine the
dimensions and structure of the balanced panel and ensure that important
variables about unit identifiers, time identifiers, and treatment and
outcome variables are correctly specified. Users can inspect this
information with the \texttt{summary()} method as well, which also
provides details about the number of unique units, time periods, and the
amount of missing data in the treatment variable. These calculations can
guide users as they consider the most appropriate way to analyze their
data and adjust matching and refinement parameters in later stages. The
\texttt{plot()} method enables users to quickly create simple figures
showing the value of a specified variable across units and time.

\begin{CodeChunk}
\begin{CodeInput}
R> dem.panel <- PanelData(panel.data = dem, 
+                        unit.id = "wbcode2", 
+                        time.id = "year", 
+                        treatment = "dem", 
+                        outcome = "y")
R> 
R> print(dem.panel)
\end{CodeInput}
\begin{CodeOutput}
Unit identifier: wbcode2
Time identifier: year
Treatment variable: dem
Outcome variable: y
Dimensions: 9384 x 5
  wbcode2 year dem  y  tradewb
1       2 1960   0 NA       NA
2       2 1961   0 NA 11.47824
3       2 1962   0 NA 12.97522
4       2 1963   0 NA 18.52119
5       2 1964   0 NA 25.75280
... [9379 more row(s) not printed]
\end{CodeOutput}
\begin{CodeInput}
R> summary(dem.panel)
\end{CodeInput}
\begin{CodeOutput}
                             quantity   value
1                             Unit ID wbcode2
2                             Time ID    year
3                           Treatment     dem
4                             Outcome       y
5                      # Unique Units     184
6               # Unique Time Periods      51
7 % of Treatment Periods Missing Data  6.9373
\end{CodeOutput}
\end{CodeChunk}

Next, users should examine the distribution of their treatment variable
using the \texttt{DisplayTreatment()} function and determine a quantity
of interest (QOI). Then, matched sets can be created using the
\texttt{PanelMatch()} function and evaluated with a set of diagnostic
functions.

The \texttt{PanelMatch()} function is responsible for both creating and
refining matched sets, described in Sections \ref{matching-process} and
\ref{Refinement}, respectively. At each of these stages, users will
specify a variety of arguments that control different aspects of the
procedures. An overview of key arguments is presented in Table
\ref{tab:pm-parameters}, but see the function documentation for full
descriptions of every parameter.

\subsection{Visualizing Treatment
Distribution}\label{visualizing-treatment-distribution}

The provided \texttt{DisplayTreatment()} function enables researchers to
visualize the distribution of the treatment variable across units and
time. This step helps users assess what analyses and comparisons can be
credibly executed, gain a general sense of how their data is structured,
and check for errors or other oddities \citep{imai2021matching}.

Using the included version of the \citet{acemoglu_democracy_2017} data
as an example, one can create a basic treatment distribution plot with
the following code.

\begin{CodeChunk}
\begin{CodeInput}
R> DisplayTreatment(panel.data = dem.panel)
\end{CodeInput}
\end{CodeChunk}

While one can create simple plots easily, some additional customization
may be desirable. For instance, user-specified labels can help clarify
the substantive interpretation of the figures and visual adjustments
might be necessary to accommodate larger data sets, as automatically
generated labels will become illegible. To this end, the
\texttt{DisplayTreatment()} function offers a large number of options
for adjusting common features of the plot. Additionally, the
\texttt{DisplayTreatment()} function returns a \texttt{ggplot2} object
(created using \texttt{geom\_tile()}), meaning that standard
\texttt{ggplot2} syntax can be used to further customize any aspect of
the figure \citep{ggplot_2}. In the code below, we take advantage of
this flexibility, applying custom labels and adding a legend.

\begin{CodeChunk}
\begin{CodeInput}
R> DisplayTreatment(panel.data = dem.panel,
+                   xlab = "Year", 
+                   ylab = "Countries", 
+                   legend.position = "bottom",
+                   legend.labels = c("Autocracy (Control)", 
+                                     "Democracy (Treatment)"),
+                   title = "Democracy as the Treatment") + 
+   scale_x_discrete(breaks = c(1960, 1970, 1980, 1990, 2000, 2010)) +
+   theme(axis.text.x = element_text(size = 8),
+         axis.text.y = element_blank(),
+         axis.ticks.y = element_blank())
\end{CodeInput}
\begin{figure}

{\centering \includegraphics{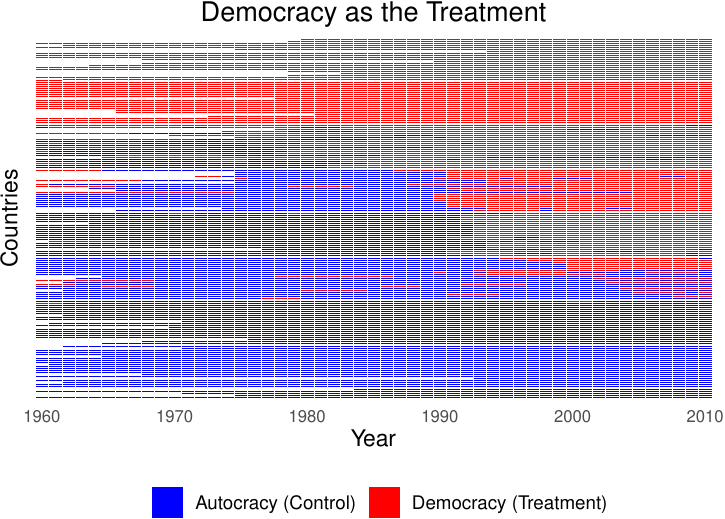} 

}

\caption{\textbf{Visualization of Treatment Distribution.} This plot shows the distribution of treatment in the data set created by Acemoglu et al. (2019), which tracks the state of democratization across all countries from 1960 to 2010. Countries are represented along the vertical axis and time is represented along the horizontal axis. Periods in which a country is classified as a democracy appear as red tiles, where as autocratic periods are shown as blue tiles. White tiles indicate missing treatment data.}\label{fig:unnamed-chunk-3}
\end{figure}
\end{CodeChunk}

In the plot, the horizontal axis represents time while the vertical axis
displays the different units in the data set. By default, red tiles
indicate periods during which a given unit is in a ``treated'' period
(i.e., the treatment variable is equal to one) and blue tiles indicate
``control'' periods (i.e., the treatment variable is equal to zero). In
this example, the ``treatment'' status corresponds to whether the
country is a democracy in a given year, as determined by
\citet{acemoglu_democracy_2017}. White spaces indicate missing treatment
variable data. Units are ordered from bottom to top in ascending order
according to the total amount of treatment ``received'' over the full
time period, with units receiving the least amount of treatment at the
bottom and the most treatment at the top.

\begin{table}[t!]
\footnotesize
    \centering
    \begin{tabular}{p{0.2\linewidth} | p{0.75\linewidth}}
      Argument Name  & Description \\ \hline
      \code{qoi} & Quantity of interest, provided as a string. Supported values are: \code{"att"}, \code{"art"}, \code{"atc"}, and \code{"ate"}. \\
      \hline
      \code{lag} & Single integer value that specifies the size of the lag window. Matched treated and control observations share an identical treatment history from \code{t-lag} to \code{t-1}, where $t$ is the time of treatment. Equivalent to $L$.\\
      \hline
      \code{lead} & Integer vector specifying the lead(s) (each entry in the vector is a value of $F$). It specifies the time periods after treatment for which point estimates and standard errors will be estimated. \\
      \hline
      \code{refinement.method} & String specifying the method for calculating weights for control units in matched sets. Supported values are: \code{"mahalanobis"} (Mahalanobis distance matching), \code{"ps.match"} (propensity score matching), \code{"CBPS.match"} (covariate balanced propensity score matching), \code{"ps.weight"} (propensity score weighting), and \code{"CBPS.weight"} (covariate balanced propensity score weighting). When \code{"mahalanobis", "ps.match",} or \code{"CBPS.match"} is specified, users may also specify \code{size.match} (see Section \ref{matching} for more.).\\
      \hline
      \code{covs.formula} & One sided formula object specifying which covariates to use for similarity calculations in the refinement procedure, separated by \code{+}. Variable transformations using standard formula syntax (e.g. \code{I(x\textasciicircum 2), I(lag(x, 1:4))}) are also supported.  \\
      \hline
    \end{tabular}
    
    \caption{\label{tab:pm-parameters} \textbf{Key PanelMatch() Arguments.} Users can adjust the matching and refinement processes by using the function arguments listed in the table. The \code{qoi}, \code{lag}, and \code{lead} arguments will affect the process of identifying treated observations and matching them to control observations using treatment history. Refinement procedures are affected primarily by the \code{refinement.method} and \code{covs.formula} arguments.}
\end{table}

\subsection{Determining the quantity of interest}\label{QOI}

After visualizing the treatment variation, users must consider the
causal quantity of interest (QOI). The package supports matching and
estimation methods for four different causal quantities of interest, via
the \texttt{qoi} argument of the \texttt{PanelMatch()} function: the
average treatment effect on the treated (ATT), the average treatment
effect of treatment reversal among the reversed (ART), the average
treatment effect on the control (ATC), and the average treatment effect
(ATE). These are specified to the \texttt{PanelMatch()} function by
setting \texttt{qoi} to \texttt{"att"}, \texttt{"art"}, \texttt{"atc"},
or \texttt{"ate"}, respectively. This article will focus on estimating
the ATT and the ART, as these tend to be of substantive interest for
many applied researchers. For users who are interested in estimating the
ATC or ATE, the package functions identically -- one need only modify
the \texttt{qoi} argument. The ATT and the ART are defined in equations
\eqref{eq:att} and \eqref{eq:art}. Appendix \ref{other-qoi} provides
definitions of the ATC and the ATE.

Researchers must also specify two key parameters when defining the
quantity of interest. The first parameter is the size of the lag window,
\textit{L}. This parameter controls for the possibility that a unit's
treatment history could act as a confounder affecting the outcome and
treatment variables and thus improves the strength of the
unconfoundedness assumption made by the method
\citep{imaikim19, imai2021matching}. Next is the number of lead periods,
\textit{F}, which specifies the number of time periods after treatment
occurs for which the quantity of interest will be estimated. These are
ultimately provided to the \texttt{PanelMatch()} function via the
\texttt{lag} and \texttt{lead} arguments, respectively. The \texttt{lag}
parameter is provided as a single integer value that defines the number
of pre-treatment periods used for exact treatment history matching. To
specify the \texttt{lead} argument, users provide one or more integers
in a vector, where each integer corresponds to a value of \(F\).
Estimates and standard errors for the specified \texttt{qoi} will then
be calculated for each value of \(F\) in the vector.

The ATT is defined formally as:

\begin{eqnarray}\label{eq:att}
  \delta_{ATT}(F,L) 
& = & \mathbb{E}\left\{Y_{i,t+F}\left(X_{it} = 1, X_{i,t-1}=0, \{X_{i,t-\ell}\}_{\ell=2}^{L}\right)\right. - \nonumber \\
& & \hspace{.3in} \left. Y_{i,t+F} \left(X_{it} = 0, X_{i,t-1}=0, \{X_{i,t-\ell}\}_{\ell=2}^{L}\right)
  \mid X_{it} = 1, X_{i,t-1} = 0\right\} \label{eq:qoi.did}
\end{eqnarray}

and the ART as:

\begin{eqnarray}\label{eq:art}
\delta_{ART}(F,L) & = &\mathbb{E}\left\{Y_{i,t+F}\left(X_{it} = 0, X_{i,t-1}=1,
  \{X_{i,t-\ell}\}_{\ell=2}^{L}\right)\right. - \nonumber \\ & & \hspace{.3in} \left. Y_{i,t+F}
  \left(X_{it} = 1, X_{i,t-1}=1, \{X_{i,t-\ell}\}_{\ell=2}^{L}\right) \mid
  X_{it} = 0, X_{i,t-1} = 1\right\} \label{eq:qoi.art}
\end{eqnarray}

where \(i\) indexes over units, \(t\) indexes over time, \(X\) is the
treatment variable and \(Y\) is the outcome variable. Intuitively, the
major difference between the ATT and the ART is as follows. The ATT
estimates the effect of receiving treatment for a unit that was under
control at time \(t-1\) on its outcome at \(t+F\) against the
counterfactual of not receiving treatment. The ART estimates the effect
of reverting to a control state after having previously received
treatment against the counterfactual of remaining in the treated state.
In the context of \citet{acemoglu_democracy_2017}, the ATT is the effect
of democratization on a country's GDP and the ART is the effect of
reverting to an autocratic regime at time \(t\) for democracies.

\subsection{Creating Matched Sets}\label{matching-process}

With the quantity of interest, lag, and lead parameters defined, the
first major procedure carried out by the \texttt{PanelMatch()} function
is the creation of matched sets of treated and control observations.
First, treated observations, \((i,t)\), are identified. The criteria for
a ``treated'' observation differs based on the specified quantity of
interest, but these are the observations for which the package will
attempt to identify a set of ``control'' units. We refer to these as
treated observations throughout this article for simplicity. Similarly,
we refer to units matched to these treated observations as control
observations.

For instance, when the ATT is the quantity of interest, the package
identifies treated observations as units that receive treatment at time
\(t\) but were in a control condition at \(t = t-1\). For the ART, the
package looks for the opposite: units that revert from a treated state
at \(t-1\) to a control state at time \(t\). Then, for each of these
identified observations, the package constructs a ``matched set'' of
control observations by identifying units that share an identical
treatment history with the treated observation from \(t-L\) to \(t-1\),
but have opposite treatment status at time \(t\).

Figure \ref{fig:toy} illustrates this process for the ATT (left panel)
and the ART (right panel), respectively. For this example, we set \(L\)
to \(3\). When the ATT is the quantity of interest, the package
identifies unit \((i,t) = (1,4)\) in the left panel of the figure as a
treated unit (circle) as the unit changed its treatment status from
\(t=3\) to \(t=4\). This unit is then matched to units \((2,4)\) and
\((4,4)\) (triangle) as they share an identical treatment status for the
three periods (rectangle) from \(t=1\) to \(t=3\) while they remained
under the control status from \(t=3\) to \(t=4\). Similarly, observation
\((5,6)\) is another treated unit, matched to units \((2,6)\) and
\((4,6)\) with identical treatment histories in the three pre-treatment
periods from \(t=3\) to \(t=5\). On the other hand, when the specified
quantity of interest is the ART, the \texttt{PanelMatch()} function
identifies observation \((i,t)=(1,5)\) as ``treated'' because it changed
its treatment status from treatment to control in \(t=5\). This unit is
then matched to observation \((3,5)\) as it remained under the treatment
condition in \(t=5\), but has an identical treatment history from
\(t=2\) to \(t=4\).

Note that there could be treated observations with empty matched sets.
For instance, observation \((3,4)\) is a treated unit when the ATT is
the quantity of interest. However, it does not have any matched control
units with an identical treatment history from \(t=1\) to \(t=3\).
Researchers should carefully examine matched set sizes in their
applications, as a large number of treated units with empty matched sets
may require them to adjust the target population.

\begin{figure}[t!]
  \spacingset{1}
  \begin{minipage}{.5\textwidth}
    \hspace{-.5in}
    \begin{tikzpicture}[->,>=stealth',shorten >=1pt,auto,node
      distance=1.1cm, thick,
      square/.style={regular polygon,regular polygon sides=4, inner
        sep=3.5},
      triangle/.style = { regular polygon, regular polygon sides=3,
        inner sep=1}]

      \tikzset{empty node/.style={circle,font=\sffamily\bfseries}}

      \tikzset{treated node/.style={circle,draw,font=\sffamily\bfseries}}

      \tikzset{matched node/.style={triangle,draw,font=\sffamily\bfseries}}

      \node[empty node] (1) {\Large$0$};
      \node[empty node] (ta) [left of=1] {$t=6$};
      \node[empty node] (tb) [below of=ta] {$t=5$};
      \node[empty node] (tc) [below of=tb] {$t=4$};
      \node[empty node] (td) [below of=tc] {$t=3$};
      \node[empty node] (te) [below of=td] {$t=2$};
      \node[empty node] (tf) [below of=te] {$t=1$};
      \node[empty node] (ua) [above of=1, yshift=-1.2ex] {$i=1$};
      \node[empty node] (ub) [right of =ua] {$i=2$};
      \node[empty node] (uc) [right of=ub] {$i=3$};
      \node[empty node] (ud) [right of=uc] {$i=4$};
      \node[empty node] (ue) [right of=ud] {$i=5$};
      \node[matched node, draw=blue, line width=0.5mm] (2) [right of=1] {\Large$0$};
      \node[empty node] (3) [right of=2] {\Large$0$};
      \node[matched node, draw=blue, line width=0.5mm, yshift=-.6ex] (4) [right of=3] {\Large$0$};
      \node[empty node, draw=blue, circle, line width=0.5mm] (5) [right of=4]{\Large$\textcolor{blue}{{\bf 1}}$};
      
      \node[empty node] (6) [below of=1]{\Large$0$};
      \node[empty node] (7) [right of=6]{\Large$0$};
      \node[empty node] (8) [right of=7]{\Large$1$};
      \node[empty node] (9) [right of=8]{\Large$0$};
      \node[empty node] (10) [right of=9]{\Large$0$};
      \node[treated node, draw=red, line width=.5mm] (11) [below of=6]{\Large$\textcolor{red}{{\bf 1}}$};
      \node[matched node, draw=red, line width=.5mm, yshift=-.6ex] (12) [right of=11]{\Large$0$};
      \node[empty node] (13) [right of=12]{\Large$1$};
      \node[matched node, draw=red, line width=.5mm, yshift=-.8ex] (14) [right of=13, yshift=.2ex]{\Large$0$};
      \node[empty node] (15) [right of=14]{\Large$0$};
      \node[empty node] (16) [below of=11]{\Large$0$};
      \node[empty node] (17) [right of=16]{\Large$0$};
      \node[empty node] (18) [right of=17]{\Large$0$};
      \node[empty node] (19) [right of=18, yshift=.2ex]{\Large$0$};
      \node[empty node] (20) [right of=19]{\Large$0$};
      \node[empty node] (21) [below of=16]{\Large$0$};
      \node[empty node] (22) [right of=21]{\Large$0$};
      \node[empty node] (23) [right of=22]{\Large$0$};
      \node[empty node] (24) [right of=23, yshift=.2ex]{\Large$0$};
      \node[empty node] (25) [right of=24]{\Large$0$};
      \node[empty node] (26) [below of=21]{\Large$0$};
      \node[empty node] (27) [right of=26]{\Large$0$};
      \node[empty node] (28) [right of=27]{\Large$1$};
      \node[empty node] (29) [right of=28, yshift=.2ex]{\Large$0$};
      \node[empty node] (30) [right of=29]{\Large$0$};
      \node[empty node] at (-2,-2.7) (21) [rotate=90]{\large{Time
          Periods}};
      \node[empty node] (22) [above of=3, yshift=2ex] {\large{Units}};
      \draw [draw=red, line width=.5mm] (-0.35,-2.9) rectangle ++(0.7,-3);
      \draw [draw=red, line width=.5mm] (0.75,-2.9) rectangle ++(0.7,-3);
      \draw [draw=red, line width=.5mm] (2.95,-2.9) rectangle ++(0.7,-3);
      \draw [draw=blue, line width=0.5mm] ($(7.north west) + (-0.2,0.1)$) rectangle ($(17.south east) + (0.2,-0.1)$);

      \draw [draw=blue, line width=0.5mm] ($(10.north west) + (-0.2,0.1)$) rectangle ($(20.south east) + (0.2,-0.1)$);

      \draw [draw=blue, line width=0.5mm] ($(9.north west) + (-0.2,0.1)$) rectangle ($(19.south east) + (0.2,-0.1)$);
    \end{tikzpicture}
    \caption*{(a) Matched Sets for ATT}
  \end{minipage}
  \begin{minipage}{.5\textwidth}
    \hspace{-.35in}
    \begin{tikzpicture}[->,>=stealth',shorten >=1pt,auto,node
      distance=1.1cm, thick,
      square/.style={regular polygon,regular polygon sides=4, inner
        sep=3.5},
      triangle/.style = { regular polygon, regular polygon sides=3,
        inner sep=1}]

      \tikzset{empty node/.style={circle,font=\sffamily\bfseries}}

      \tikzset{treated node/.style={circle,draw,font=\sffamily\bfseries}}

      \tikzset{matched node/.style={triangle,draw,font=\sffamily\bfseries}}
      
      \draw [draw=red, line width=.5mm] (-0.35,-1.75) rectangle ++(0.7,-3);
      \draw [draw=red, line width=.5mm] (1.85,-1.75) rectangle ++(0.7,-3);

      
      \node[empty node] (1) {\Large$0$};
      \node[empty node] (ta) [left of=1] {$t=6$};
      \node[empty node] (tb) [below of=ta] {$t=5$};
      \node[empty node] (tc) [below of=tb] {$t=4$};
      \node[empty node] (td) [below of=tc] {$t=3$};
      \node[empty node] (te) [below of=td] {$t=2$};
      \node[empty node] (tf) [below of=te] {$t=1$};
      \node[empty node] (ua) [above of=1, yshift=-1.2ex] {$i=1$};
      \node[empty node] (ub) [right of =ua] {$i=2$};
      \node[empty node] (uc) [right of=ub] {$i=3$};
      \node[empty node] (ud) [right of=uc] {$i=4$};
      \node[empty node] (ue) [right of=ud] {$i=5$};
      \node[empty node] (2) [right of=1] {\Large$0$};
      \node[empty node] (3) [right of=2] {\Large$0$};
      \node[empty node] (4) [right of=3, yshift=.2ex]{\Large$0$};
      \node[empty node] (5) [right of=4]{\Large$1$};
      \node[treated node, draw=red, line width=.5mm] (6) [below of=1]{\Large$\textcolor{red}{{\bf 0}}$};
      \node[empty node] (7) [right of=6]{\Large$0$};
      \node[matched node, draw=red, line width=.5mm] (8) [right of=7]{\Large$1$};
      \node[empty node] (9) [right of=8]{\Large$0$};
      \node[empty node] (10) [right of=9]{\Large$0$};
      \node[empty node] (11) [below of=6]{\Large$1$};
      \node[empty node] (12) [right of=11]{\Large$0$};
      \node[empty node] (13) [right of=12]{\Large$1$};
      \node[empty node] (14) [right of=13, yshift=.2ex]{\Large$0$};
      \node[empty node] (15) [right of=14]{\Large$0$};
      \node[empty node] (16) [below of=11]{\Large$0$};
      \node[empty node] (17) [right of=16]{\Large$0$};
      \node[empty node] (18) [right of=17]{\Large$0$};
      \node[empty node] (19) [right of=18, yshift=.2ex]{\Large$0$};
      \node[empty node] (20) [right of=19]{\Large$0$};
      \node[empty node] (21) [below of=16]{\Large$0$};
      \node[empty node] (22) [right of=21]{\Large$0$};
      \node[empty node] (23) [right of=22]{\Large$0$};
      \node[empty node] (24) [right of=23, yshift=.2ex]{\Large$0$};
      \node[empty node] (25) [right of=24]{\Large$0$};
      \node[empty node] (26) [below of=21]{\Large$0$};
      \node[empty node] (27) [right of=26]{\Large$0$};
      \node[empty node] (28) [right of=27]{\Large$1$};
      \node[empty node] (29) [right of=28, yshift=.2ex]{\Large$0$};
      \node[empty node] (30) [right of=29]{\Large$0$};
      \node[empty node] at (-2,-2.7) (21) [rotate=90]{\large{Time
          Periods}};
      \node[empty node] (22) [above of=3, yshift=2ex] {\large{Units}};
    \end{tikzpicture}
    \caption*{(b) Matched Sets for ART}
  \end{minipage}
  \caption{{\bf An Example of Matched Sets with Five Units and Six
      Time Periods}. Panels (a) and (b) illustrate how matched sets
    are chosen for the ATT and the ART, respectively, when $L=3$. For each treated
    observation (colored circles), we select a set of control
    observations from other units in the same time period (triangles
    with the same color) that have an identical treatment history
    (rectangles with the same color).} \label{fig:toy}
\end{figure}

Matched sets for the ATT can be formally expressed as: \begin{equation}
  \mathcal{M}_{it}^{\textrm{ATT}} \ = \ \{i^\prime : i^\prime \ne i, X_{i^\prime t} =0,
                 X_{i^\prime t^\prime} = X_{i t^\prime} \textrm{ for
                 all } t^\prime = t-1, \dots,t-L\}, \label{eq:matched.set}
\end{equation} for the treated observations with \(X_{it}=1\) and
\(X_{i,t-1}=0\). A similar expression can be defined for the ART:
\begin{equation}
\mathcal{M}_{it}^{\textrm{ART}} = \{i^\prime : i^\prime \ne i, X_{i^\prime t} =1,
X_{i^\prime t^\prime} = X_{i t^\prime}, \textrm{ for all } t^\prime =
t-1, \dots,t-L\}
\end{equation} for the treated observations with \(X_{it}=0\) and
\(X_{i,t-1}=1\). See Appendix \ref{atc-msets} for a description of
matched sets for the ATC and ATE.

Larger values for the \texttt{lag} parameter will help users address
potential confounding as well as carryover effects due to past
treatments up to \(L\) periods. However, researchers should be cautious
about having many empty matched sets, as larger values of \(L\) will put
more constraints on identifying matched units. Furthermore, smaller
numbers of control observations that are matched to each treated unit
will possibly lower the precision of effect estimates. Hence,
researchers should determine appropriate values for the \texttt{lag}
parameter based on their substantive knowledge.

Figure \ref{fig:lagcomparison} shows how different \texttt{lag} values
can affect the size and number of matched sets. In the left hand panel
of Figure \ref{fig:lagcomparison}, a \texttt{lag} value of 4 is
specified in a \texttt{PanelMatch()} configuration. On the right hand
side of the figure, a \texttt{lag} value of 10 is specified, and no
other parameter adjustments are made. The number of treated observations
with no matched controls (i.e., the number of ``empty matched sets'') is
shown with a vertical red line. Under the less restrictive specification
when \texttt{lag\ =\ 4}, there are fewer empty matched sets and a larger
number of matched sets of size 60 and larger, relative to the more
restrictive configuration. Note that these distribution figures are
created using built in methods, discussed in Section \ref{mset-methods}.

\begin{CodeChunk}
\begin{figure}[t]

{\centering \includegraphics{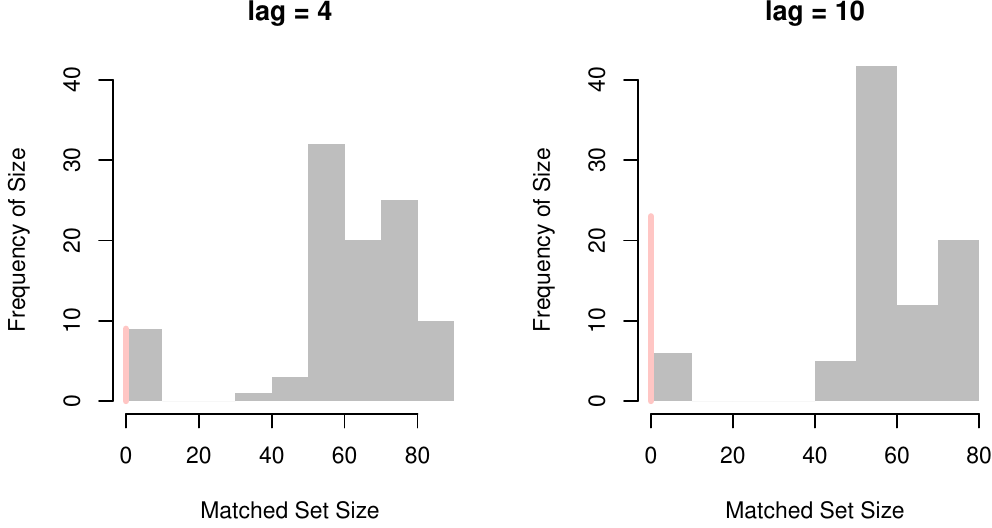} 

}

\caption{\textbf{Effects of Adjusting Lag Window on Matched Set Sizes.} This figure shows the distribution of matched set sizes in two different \texttt{PanelMatch()} configurations that differ only in the size of the lag window, \textit{L}. The number of treated observations with no matched control units are shown with a vertical red bar. The left hand plot uses a smaller lag window, with \textit{L} = 4, while the right uses a more restrictive, larger window with \textit{L = 10}. The smaller lag window results in fewer empty matched sets and a larger number of matched sets of size 60 and greater, relative to the configuration with a larger lag window.}\label{fig:lagcomparison}
\end{figure}
\end{CodeChunk}

\subsubsection{Other Matching
Parameters}\label{other-matching-parameters}

In addition to \texttt{qoi}, \texttt{lag}, and \texttt{lead}, the
\texttt{PanelMatch()} function accepts a number of other arguments that
affect the matching process. Of these,
\texttt{forbid.treatment.reversal} allows users to impose the
``staggered adoption'' assumption within the lead window. That is, this
argument, provided as a logical value, specifies whether or not treated
observations are allowed to revert to control status within the lead
window (i.e., from \(t+0\) to \(t+F\)). The default is \texttt{FALSE.}
When set to \texttt{TRUE,} only units that have treated status from time
\(t\) to \(t+F\) are used as viable treated observations. Note that,
when this option is \texttt{TRUE}, the package also requires viable
treated units to not be missing any treatment variable data over the
lead
window.\footnote{Of course, by definition, these units also have control status at time $t-1$.}
Note that we use the terminology for the ATT here for the sake of
exposition. When estimating the ART, the same rules apply, except
control status is enforced over the lead window, rather than treated
status. The \texttt{match.missing} parameter also affects the matching
procedure. Setting \texttt{match.missing\ =\ FALSE} will drop the
observations with missingness in the treatment variable in the lag
window. On the other hand, setting \texttt{match.missing\ =\ TRUE} will
keep those observations while matching on the missingness pattern in the
treatment variable.

Other arguments, described fully in the function documentation, allow
users to add or remove further constraints on how matched sets are
constructed. We generally recommend that users keep the default
specifications for these parameters unless there are clear substantive
reasons to do otherwise.

\subsubsection{Understanding PanelMatch Objects}\label{mset-methods}

The \texttt{PanelMatch()} function returns a \texttt{PanelMatch} object,
which is implemented as a list containing a \texttt{matched.set} object,
along with some additional attributes tracking metadata about the
configuration of the corresponding \texttt{PanelMatch()} call. The
package includes a number of methods to facilitate working with
\texttt{PanelMatch} objects, including functions for summarizing,
printing, and plotting.

The \texttt{print()} and \texttt{summary()} methods are highlighted in
the following code chunk. After creating a basic \texttt{PanelMatch}
object, users can leverage these methods to quickly inspect the results.
The \texttt{print()} method provides information about important
parameters, like the specified lag, lead, and QOI values, along with a
preview of information about matched sets. The \texttt{summary()} method
complements this by providing summary information about the distribution
of the sizes of matched sets, along with details about the number of
total treated units and empty matched sets. Specifically, it shows the
size of the smallest matched set, the size of the largest matched set,
the mean matched set size, the median matched set size, and the first
and third quartile sizes.

\begin{CodeChunk}
\begin{CodeInput}
R> pm.results <- PanelMatch(panel.data = dem.panel,
+                          lag = 4,
+                          refinement.method = "none",
+                          match.missing = FALSE,
+                          qoi = "att",
+                          lead = 0:2)
R> print(pm.results)
\end{CodeInput}
\begin{CodeOutput}
PanelMatch Object Summary
---------------------------------------- 
Unit ID             : wbcode2
Time ID             : year
Outcome Variable    : y
Treatment Variable  : dem
Lag                 : 4
Max Lead            : 2
Refinement Method   : none
---------------------------------------- 
QOI: ATT
---------------------------------------- 
  wbcode2 year matched.set.size
1       4 1992               74
2       4 1997                2
3       6 1973               63
4       6 1983               73
5       7 1991               81
... [104 more matched set(s) not printed]
\end{CodeOutput}
\begin{CodeInput}
R> summary(pm.results)
\end{CodeInput}
\begin{CodeOutput}
$att
                      quantity     value
1                         Min.   0.00000
2                      1st Qu.  57.00000
3                       Median  61.00000
4                         Mean  55.25688
5                      3rd Qu.  74.00000
6                         Max.  81.00000
7      Number of treated units 109.00000
8 Number of empty matched sets   9.00000
\end{CodeOutput}
\end{CodeChunk}

Additionally, users can visualize the distribution of the sizes of
matched sets using the \texttt{plot()} method. By default, a vertical
red line will indicate the number of matched sets where treated units
were unable to be matched with any control units (i.e., the number of
empty matched sets). These plots can be adjusted using the arguments
described in the documentation, or any other arguments normally passed
to the \texttt{graphics::plot()} function. If \texttt{qoi\ =\ "ate"},
then the \texttt{plot()} method will return a figure for the ATT and
ATC, in that order.

Researchers can use this information to make informed assessments about
viable analyses. For instance, having many treated observations with no
matched controls might modify the target population, and hence make it
difficult to credibly draw causal inferences. Moreover, a large number
of small matched sets generally leads to larger standard errors.

The following code shows how users can examine various diagnostics using
the \texttt{plot()} method. From Figure \ref{fig:diagnostics-methods},
we see that there are relatively few empty matched sets -- this quantity
is indicated with a vertical red line on the plot. We also observe that
there are a reasonable number of non-empty matched sets, and the sizes
of these matched sets are suitably large to proceed with additional
analyses.

\begin{CodeChunk}
\begin{CodeInput}
R> pm.results <- PanelMatch(panel.data = dem.panel,
+                          lag = 4,
+                          refinement.method = "none",
+                          match.missing = FALSE,
+                          qoi = "att",
+                          lead = 0:2)
R> plot(pm.results, xlim = c(0, 100))
\end{CodeInput}
\begin{figure}

{\centering \includegraphics{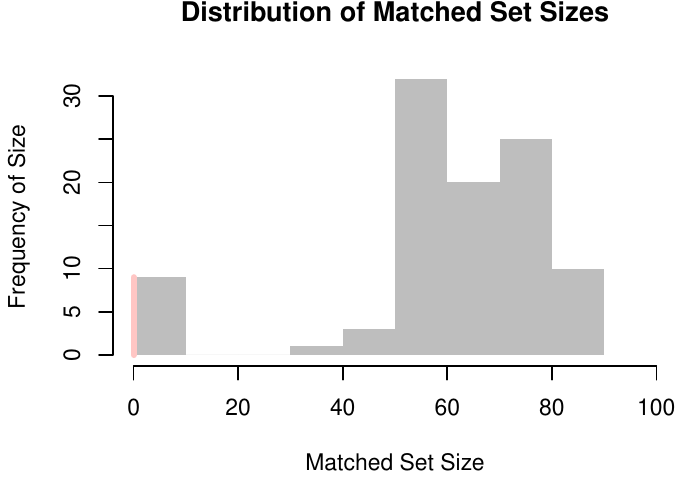} 

}

\caption{\textbf{Using the \texttt{plot()} Method to Visualize the Distribution of Matched Set Sizes.} Users can provide a \texttt{PanelMatch} object to a provided \texttt{plot()} method in order to  visualize the distribution of matched set sizes produced by their specified \texttt{PanelMatch()} configuration in a histogram. The number of empty matched sets is shown with a vertical red bar. Researchers should assess whether their matched sets are sufficiently large, as well as if they have an adequate number of non-empty matched sets.}\label{fig:diagnostics-methods}
\end{figure}
\end{CodeChunk}

\subsubsection{Understanding matched.set
Objects}\label{understanding-matched.set-objects}

Within the \texttt{PanelMatch} object, the attached \texttt{matched.set}
object is always named either \texttt{att}, \texttt{art}, or
\texttt{atc}, corresponding to the specified
\texttt{qoi}.\footnote{When \code{qoi = "ate"}, there are two \code{matched.set} objects included in the results of the \code{PanelMatch()} call.
Specifically, there will be two matched sets, named \code{att} and \code{atc}. 
This is because in practice, estimates for the ATE are calculated using a weighted average of ATT and ATC estimates.}
Users can extract the \texttt{matched.set} object affiliated with a
\texttt{PanelMatch} object(s) using the \texttt{extract()} method:

\begin{CodeChunk}
\begin{CodeInput}
R> msets <- extract(pm.results)
\end{CodeInput}
\end{CodeChunk}

The \texttt{matched.set} object is also a list with some additional
attributes. Each entry in this list corresponds to a matched set of
treated and control units and has a name that uniquely identifies that
entry. Specifically, each element has a name that is structured in the
following way:
\texttt{{[}unit\ id\ of\ treated\ unit{]}}.\texttt{{[}time\ of\ treatment{]}}.
Each element in the list is a vector indicating the control units (as a
vector of the unit identifiers) that are matched with the treated unit
specified in the name of that element. The length of this vector is
equivalent to the size of the matched set, i.e., \(|\mathcal{M}_{it}|\).
The \texttt{{[}} and \texttt{{[}{[}} operators are implemented to behave
intuitively as they would for list objects.

Methods for printing, plotting, and summarizing these objects are also
implemented. The \texttt{summary()} method calculates information about
the sizes of matched sets and the \texttt{print()} method displays this
data. The \texttt{print()} method offers a quick way to preview
information about the sizes of matched sets and treated observations.
Users can adjust the \texttt{n} and \texttt{show.all} parameters to
change how much information they want to see. The parameter \texttt{n}
adjusts the number of matched sets to print information about.
Alternatively, users can print out information about all matched sets by
setting \texttt{show.all\ =\ TRUE}.

\begin{CodeChunk}
\begin{CodeInput}
R> print(msets)
\end{CodeInput}
\begin{CodeOutput}
  wbcode2 year matched.set.size
1       4 1992               74
2       4 1997                2
3       6 1973               63
4       6 1983               73
5       7 1991               81
... [104 more matched set(s) not printed]
\end{CodeOutput}
\begin{CodeInput}
R> print(msets, n = 10)
\end{CodeInput}
\begin{CodeOutput}
   wbcode2 year matched.set.size
1        4 1992               74
2        4 1997                2
3        6 1973               63
4        6 1983               73
5        7 1991               81
6        7 1998                1
7       12 1992               74
8       13 2003               58
9       15 1991               81
10      16 1977               63
... [99 more matched set(s) not printed]
\end{CodeOutput}
\end{CodeChunk}

In contrast to the \texttt{summary()} and \texttt{print()} methods for
\texttt{PanelMatch} objects, which offer higher level information about
the distribution of matched set sizes, the output of the
\texttt{matched.set} methods is more detailed and verbose. These
\texttt{matched.set} methods will return details about treated
observations and the sizes of all matched sets, not just an overview,
primarily making them useful to advanced users of the package. See
Appendix \ref{mset-other-methods} for more on these methods.

Each control unit in each matched set also has an associated weight,
which is determined by the refinement procedure, as described in Section
\ref{Refinement}. Similarly, for some methods, the distance between
control units and their matched treated units might also be calculated.
Researchers can inspect this information with the \texttt{weights()} and
\texttt{distances()} methods for \texttt{matched.set} objects. Both of
these methods return a list of named vectors. Each vector represents a
matched set of controls. The names of the vector elements indicate the
ID of a particular matched control, and each vector element represents
the weight or distance assigned to that particular control in that
particular matched set. One can also use the subset operators to inspect
the weights or distances of a specific matched set.

In the following code, we extract the weights of the controls in the
matched sets. We then examine the matched set associated with
Afghanistan (identified by \texttt{wbcode2\ =\ 4}), which democratized
in 1992. There are 74 matched control units. In this case, all 74
control units have equal weights of \(0.01351351 = 1/74\) because no
refinement method was specified (i.e.,
\texttt{refinement.method\ =\ "none"})

\begin{CodeChunk}
\begin{CodeInput}
R> wts <- weights(msets)
R> wts[["4.1992"]]
\end{CodeInput}
\begin{CodeOutput}
         3         13         16         19         28         29         31 
0.01351351 0.01351351 0.01351351 0.01351351 0.01351351 0.01351351 0.01351351 
        35         36         37         43         45         47         51 
0.01351351 0.01351351 0.01351351 0.01351351 0.01351351 0.01351351 0.01351351 
        53         57         62         64         65         67         70 
0.01351351 0.01351351 0.01351351 0.01351351 0.01351351 0.01351351 0.01351351 
        71         81         84         87         93         95         96 
0.01351351 0.01351351 0.01351351 0.01351351 0.01351351 0.01351351 0.01351351 
        97        103        104        105        109        110        112 
0.01351351 0.01351351 0.01351351 0.01351351 0.01351351 0.01351351 0.01351351 
       114        115        116        118        123        124        125 
0.01351351 0.01351351 0.01351351 0.01351351 0.01351351 0.01351351 0.01351351 
       128        129        134        140        142        150        155 
0.01351351 0.01351351 0.01351351 0.01351351 0.01351351 0.01351351 0.01351351 
       156        157        159        161        163        168        171 
0.01351351 0.01351351 0.01351351 0.01351351 0.01351351 0.01351351 0.01351351 
       172        173        175        176        178        179        180 
0.01351351 0.01351351 0.01351351 0.01351351 0.01351351 0.01351351 0.01351351 
       182        184        186        187        190        193        196 
0.01351351 0.01351351 0.01351351 0.01351351 0.01351351 0.01351351 0.01351351 
       197        199        200        202 
0.01351351 0.01351351 0.01351351 0.01351351 
\end{CodeOutput}
\end{CodeChunk}

The package also includes a \texttt{plot()} method for
\texttt{matched.set} objects. This is discussed in Section \ref{diag}.

\section{Refining Matched Sets}\label{Refinement}

The second step executed by the \texttt{PanelMatch()} function
calculates weights for control units in each matched set. This
refinement step is intended to identify control units that are similar
to each treated unit in terms of their pre-treatment observed
characteristics. This step also helps researchers evaluate the validity
of the parallel trends assumption conditional on observed time-varying
confounders. Users will specify a set of pre-treatment covariates and a
method of calculating similarities between treated and control units for
a given matched set using these variables. They are specified using the
\texttt{covs.formula} and \texttt{refinement.method} parameters,
respectively. Then, for each matched set, the package will calculate
weights for each control unit, based on the results of the similarity
calculations and a number of other additional user specified parameters.

Researchers can choose to refine their matched sets using matching
methods based on Mahalanobis distance and propensity scores, including
those based on the covariate balancing propensity scores (CBPS)
described in \citet{imairatk14}. They are specified with
\texttt{refinement.method\ =\ "mahalanobis"},
\texttt{refinement.method\ =\ "ps.match"}, or
\texttt{refinement.method\ =\ "CBPS.match"}, respectively.

Alternatively, they can refine via weighting methods, which assign
weights to all control units in a matched set over a continuous spectrum
based on the inverse of estimated propensity scores. Valid inputs
include \texttt{refinement.method\ =\ "ps.weight"} and
\texttt{refinement.method\ =\ "CBPS.weight"}. Note that matching methods
are special cases of weighting methods. This is because they assign
equal weights to a subset of control units in each matched set that meet
or exceed a threshold of similarity to their matched treated unit.

\subsection{Refinement with Matching Methods}\label{matching}

Matching methods refine matched sets in the following way. First, the
package calculates pairwise similarities between treated and control
units in a matched set, based on a distance measure of choice. Formally,
we calculate \(S_{it}(i^\prime)\) for all
\(i^\prime \in \mathcal{M}_{it}\). Then, a (sub)set of control units
most similar to the matched treated unit is identified and assigned
equal, non-zero weights. When using a matching method for refinement,
users may calculate these pairwise distances based on Mahalanobis
distance (\texttt{refinement.method\ =\ "mahalanobis"}), or propensity
scores (\texttt{refinement.method\ =\ "ps.match"} or
\texttt{refinement.method\ =\ "CBPS.match"}).

When the user specifies a Mahalanobis distance based refinement process,
the package computes a standardized Mahalanobis distance between every
control unit in a matched set and the corresponding treated unit using
the specified covariates and averages it across time periods. Formally,
the package calculates the pairwise distance between treated and a given
matched control unit, \(i^\prime \in \mathcal{M}_{it}\) according to the
following formula: \begin{equation}
  S_{it}(i^\prime) \ = \ \frac{1}{L} \sum_{\ell= 1}^{L} \sqrt{(\mathbf{V}_{i,t-\ell} -
  \mathbf{V}_{i^\prime, t-\ell})^\top \boldsymbol{\Sigma}_{i,t-\ell}^{-1} (\mathbf{V}_{i,t-\ell} -
  \mathbf{V}_{i^\prime, t-\ell}) } \label{eq:maha}
\end{equation} where \(\mathbf{V}_{it^\prime}\) is a matrix containing
the user-specified, time-varying covariates one wishes to adjust for and
\(\boldsymbol{\Sigma}_{it^\prime}\) is the sample covariance matrix of
\(\mathbf{V}_{i t^\prime}\).

If a refinement method based on propensity scores is specified,
propensity scores are first estimated using logistic regression or the
CBPS. Then, given propensity score estimates for each unit, the
following distance calculation is used for each control unit in a
matched set: \begin{equation}
  S_{it}(i^\prime) \ = \ |{\rm
    logit}\{\hat{e}_{it}(\{\mathbf{U}_{i,t-\ell}\}_{\ell = 1}^{L})\} -
  {\rm logit}\{\hat{e}_{i^\prime t}(\{\mathbf{U}_{i^\prime,
    t-\ell}\}_{\ell = 1}^{L})\}| \label{eq:pscore.distance}
\end{equation} where \(i^\prime \in \mathcal{M}_{it}\) and
\(\hat{e}_{i^\prime t}(\{\mathbf{U}_{i,t-\ell}\}_{\ell = 1}^{L})\) is
the estimated propensity score. We define
\(\mathbf{U}_{it^\prime} = (X_{it^\prime}, \mathbf{V}_{it^\prime}^\top)^\top\),
where, as before, \(\mathbf{V}_{it^\prime}\) represents the
user-specified, time-varying covariates one wishes to adjust for.

The set of pre-treatment covariates that are used in all refinement
calculations should be specified via the \texttt{covs.formula} argument.
The \texttt{covs.formula} argument is specified as a one-sided formula
object, with the names of the desired pre-treatment variables separated
by \texttt{+}. For example, to include variables \texttt{x} and
\texttt{w} for use in refinement, one would specify
\texttt{covs.formula\ =\ \textasciitilde{}\ x\ +\ w}. These variables
will be used during the refinement process to calculate similarities
between treated and control units, which are then used to determine the
weights assigned to control units.

The package allows for users to apply standard variable transformations
as well, so inputs like \code{I(x\textasciicircum 2)} are valid. To that
end, users may also wish to include ``lagged'' control variables in the
\texttt{covs.formula} specification. To do this, they will utilize the
included \texttt{lag()} function, which accepts the name of the variable
users wish to lag as the first argument, and the number of lags to
calculate as a vector in the second argument. So, for instance, if a
user wished to include 4 lags of the \texttt{x} control variable such
that values of x at \(t-1, t-2, t-3\), and \(t-4\) are included as
control variables, it would be done in the following way:
\texttt{\textasciitilde{}I(lag(x,\ 1:4))}. As a general suggestion, if
users want to make the parallel trend assumption, then do not include
lagged dependent variables in covariate refinement. However, if lagged
outcomes are assumed to be confounders, affecting both the outcome and
treatment, then one should include them.

After calculating these distances, \(S_{it}(i^\prime)\) for all
\(i^\prime \in \mathcal{M}_{it}\), all matching-based refinement methods
will identify a set of control units most similar to each treated unit
and assign these control units within a matched set identical, non-zero
weights. The number of control units assigned non-zero weights is
determined by an additional \texttt{PanelMatch()} argument,
\texttt{size.match}. This argument is specified by users as some
integer, \(N\). After calculating the pairwise distances for each
matched set, the \(N\)-th smallest distance is determined. That is, we
calculate \(S^{(N)}_{it}\), where \(S^{(N)}_{it}\) is the \(N\)th-order
statistic of the pairwise distances calculated for a given matched set.
All matched control units that are less than or equal to this distance
away from the treated unit receive a weight of \(1/N\). All other
control units in the matched set receive a weight of
0.\footnote{In practice, it is possible that the number of control units with non-zero weight might exceed \texttt{size.match}.
This could occur when there are a number of control units that have the exact same similarity score.
For instance, say there are 10 control units matched to a treated unit, \texttt{set.size = 5}, but there are 8 control units that are exactly the same distance away from the treated unit.
These types of cases are handled by including units that are less than or equal to the $set.size$-th smallest distance.
So, 8 units would receive a weight of 1/8 in this example.
When distances are calculated using continuous data, these situations are rare.}

The following code provides an example of a \texttt{PanelMatch()}
specification using a matching method for refinement. Specifically, it
uses Mahalanobis distance based refinement. However, propensity score
based methods are also available, as described previously.

\begin{CodeChunk}
\begin{CodeInput}
R> PM.maha <- PanelMatch(panel.data = dem.panel, 
+                       lag = 4, 
+                       refinement.method = "mahalanobis",
+                       match.missing = FALSE, 
+                       covs.formula = ~ I(lag(tradewb, 0:4)) + 
+                                        I(lag(y, 1:4)),
+                       size.match = 5, 
+                       qoi = "att", 
+                       lead = 0:2,
+                       use.diagonal.variance.matrix = TRUE,
+                       forbid.treatment.reversal = FALSE)
\end{CodeInput}
\end{CodeChunk}

\subsection{Refinement with Weighting Methods}\label{weighting}

Weighting methods assign a set of normalized weights to every control
unit \(i^\prime\) in a given matched set such that units more similar to
the treated observation \((i,t)\) are given higher weights and less
similar units are given lower weights. The package relies on inverse
propensity score weighting methods in order to calculate these weights
\citep{hirano}. Users will specify whether they want these weights
calculated using propensity scores from a logistic regression or those
generated by the CBPS method. Formally, the weight,
\(w_{it}^{i^\prime}\), for a control unit, \(i^\prime\), within a given
matched set for treated observation \((i,t)\) are calculated with the
following: \begin{equation}
  w_{it}^{i^\prime} \ \propto \ \frac{\hat{e}_{i^\prime
      t}(\{\mathbf{U}_{i,t-\ell}\}_{\ell = 1}^{L})}{1- \hat{e}_{i^\prime t}(\{\mathbf{U}_{i,t-\ell}\}_{\ell =
    1}^{L})} \label{eq:prop.weight}
\end{equation} such that
\(\sum_{i^\prime \in \mathcal{M}_{it}} w_{it}^{i^\prime} = 1\) and
\(w_{it}^{i^\prime} = 0\) for \(i^\prime \notin \mathcal{M}_{it}\). Each
\(\hat{e}_{i^\prime t}\) is an estimated propensity score, either via
logistic regression or CBPS.

As with matching methods, the method of refinement and pre-treatment
covariates to be used for similarity calculations are specified to the
\texttt{refinement.method} and \texttt{covs.formula} arguments,
respectively. Valid weighting refinement methods include
\texttt{"ps.weight"} and \texttt{"CBPS.weight"}. The following code
shows an example of a \texttt{PanelMatch()} configuration doing matched
set refinement with a weighting method based on propensity scores.

\begin{CodeChunk}
\begin{CodeInput}
R> PM.ps.weight <- PanelMatch(lag = 4, 
+                       refinement.method = "ps.weight",
+                       panel.data = dem.panel, 
+                       match.missing = FALSE, 
+                       covs.formula = ~ I(lag(tradewb, 0:4)) + 
+                                        I(lag(y, 1:4)),
+                       qoi = "att", 
+                       lead = 0:2,
+                       use.diagonal.variance.matrix = TRUE,
+                       forbid.treatment.reversal = FALSE)
\end{CodeInput}
\end{CodeChunk}

Figure \ref{fig:refinement-vis} visualizes the differences between
matching and weighting methods on a single matched set. The plot on the
left shows the treatment variation plot for units in a single matched
set before any refinement is applied. All units have equal weights and
share an identical treatment history from \(t-L\) to \(t-1\), where
\(L = 4\). The center plot, with darker colors on the plot indicating
higher weights, shows the effects of a weighting method. Control units
are assigned a range of non-zero weights. Shown in the rightmost plot, a
matching method will effectively create a subset of the original matched
set, with some number of the most similar control units all receiving
equal, non-zero weights.

\begin{CodeChunk}
\begin{figure}[t]

{\centering \includegraphics{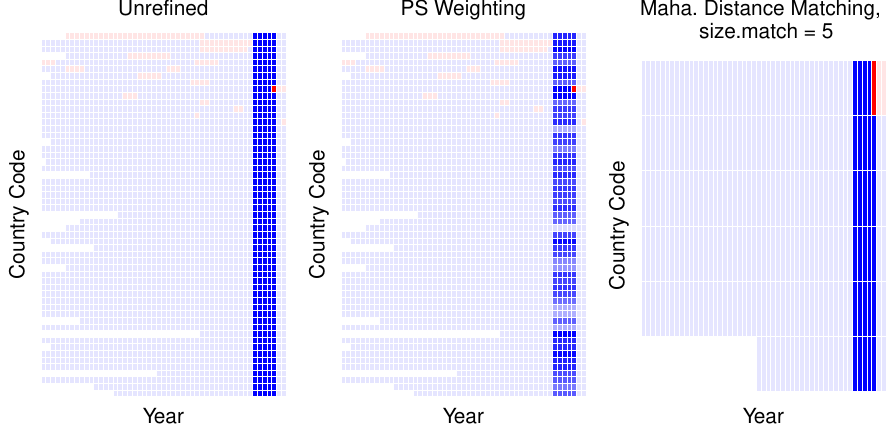} 

}

\caption{\textbf{Visualizing the Differences Between Weighting and Matching Refinement Methods.} The first plot on the left shows the treatment history of a single treated observation and all matched controls, highlighting the lag period ($L = 4$). All control observations have equal weight. The middle plot uses a gradient to visualize the weights of control units in the same matched set after the matched set has been refined using a propensity score weighting method. Control units more similar to the treated unit receive higher weights and have darker colors. The last plot shows the effects of refinement using a matching method: setting the \texttt{size.match} parameter to 5, the five most similar matched control observations receive equal, non-zero weights. Such plots can be generated by providing inputs to the \texttt{matched.set}, \texttt{show.set.only} and \texttt{gradient.weights} arguments of the \texttt{DisplayTreatment()} function. See the function documentation in the package for a complete description of all \texttt{DisplayTreatment()} arguments.}\label{fig:refinement-vis}
\end{figure}
\end{CodeChunk}

\subsection{Refinement Diagnostics}\label{diag}

As discussed in Section \ref{Refinement}, the refinement procedures are
intended to help control for pre-treatment confounding variables and
meet the parallel trend assumption more credibly. To this end, users can
explore potential sources of bias and examine which units in the data
are generally being included and/or receiving weights from matched sets
with the \texttt{plot()} method for \texttt{matched.set} objects. Users
should inspect these results to see if units are being unexpectedly
excluded or overused.

An example figure can be seen in Figure \ref{fig:plotms-vis}, examining
the results of the propensity score based refinement configuration
specified previously. The y-axis indexes treated observations (i.e.,
\((i,t)\) pairs), and matched control units are plotted along the
x-axis. Tiles are colored according to the weight a particular unit
received in a matched set. Users should inspect these results to see if
units are being unexpectedly excluded or overly relied upon for weights.
From the figure, we can observe that some units are used many times as
matched controls, while a sizable number are never used at all. Users
should apply their substantive knowledge at this stage to see if any
observed patterns are cause for concern.

\begin{CodeChunk}
\begin{CodeInput}
R> ps.weight.msets <- extract(PM.ps.weight)
R> plot(ps.weight.msets, panel.data = dem.panel)
\end{CodeInput}
\end{CodeChunk}

\begin{CodeChunk}
\begin{figure}[t]

{\centering \includegraphics{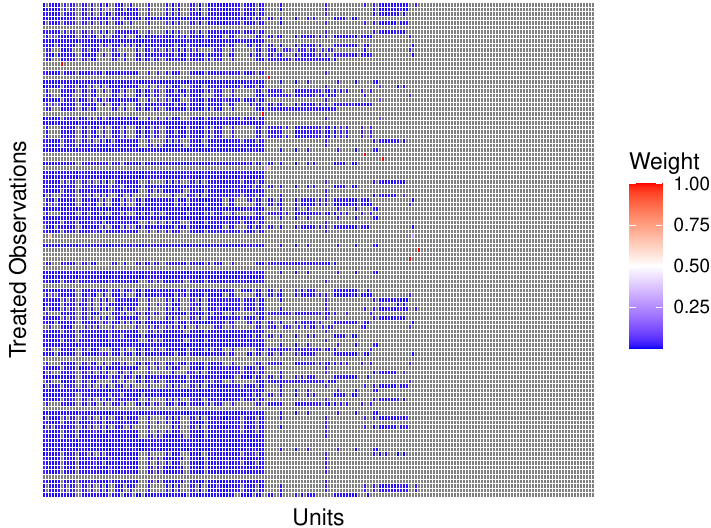} 

}

\caption{\textbf{Using the \texttt{plot()} Method for \texttt{matched.set}  Objects}. Some units are used many times as matched controls, while others are never used. Users should consider whether or not patterns like these are substantively meaningful.}\label{fig:plotms-vis}
\end{figure}
\end{CodeChunk}

Researchers can assess the impact of refinement by calculating covariate
balance. Specifically, the package implements the method suggested by
\citet{imai2021matching} that calculates a standardized mean difference
between each matched treated and control unit for all specified
covariates for each pre-treatment period from \texttt{t-lag} to
\texttt{t-1}. This is then averaged across all matched sets.

Formally, for a given treated observation \((i,t)\), this balance metric
is defined as: \begin{equation} 
B_{it}(j, \ell) \ = \ \frac{ V_{i,t-\ell, j} -
\sum_{i^\prime \in \mathcal{M}_{it}} w_{it}^{i^\prime} \ V_{i^\prime, t-\ell,
j} }{\sqrt{\frac{1}{N_1 -1}\sum_{i^\prime = 1}^{N}
\sum_{t^\prime=L+1}^{T-F} D_{i^\prime t^\prime} (V_{i^\prime,
t^\prime-\ell, j} - \overline{V}_{t^\prime-\ell, j})^2}}
\end{equation} where \(j\) indexes over covariates (so \(V_{i,t, j}\) is
the \(j\)th variable of \(\mathbf{V}\), for unit \(i\) at time \(t\)),
\(t-\ell\) indicates the pre-treatment time period, and
\(D_{it}=X_{it} (1-X_{i,t-1})\cdot\mathbf{1}\{|\mathcal{M}_{it}| > 0 \}\).
Note that this definition is specific to the ATT. For the ART, it is
defined as
\(D_{it}=(1-X_{it}) X_{i,t-1}\cdot\mathbf{1}\{|\mathcal{M}_{it}| > 0 \}\).
In other words, for the ATT (ART), \(D_{it} = 1\) if observation
\((i,t)\) changes to the treatment (control) status at time \(t\) from
the control (treatment) status at time \(t-1\) and has at least one
matched control unit. Thus,
\(N_1 = \sum_{i^\prime =1}^N \sum_{t^\prime=L+1}^{T-F} D_{i^\prime t^\prime}\)
is the total number of treated observations and
\(\overline{V}_{t-\ell,j}=\sum_{i=1}^N V_{i,t-\ell,j}/N\).

Aggregating this covariate balance measure across all treated
observations for each covariate and pre-treatment time period gives the
following: \begin{eqnarray}
  \overline{B}(j, \ell) & = & \frac{1}{N_1} \sum_{i=1}^N
                              \sum_{t=L+1}^{T-F} D_{it} B_{it}(j,
                              \ell) \label{eq:balance}
\end{eqnarray}

We can now investigate the effectiveness of different matching and
refinement methods using the \texttt{get\_covariate\_balance()}
function. This function calculates the covariate balance measure
\(\overline{B}(j, \ell)\) defined in equation \eqref{eq:balance} for
each specified covariate from time \texttt{t-lag} to \texttt{t-1}.
Researchers should hope to see small values of
\(\overline{B}(j, \ell)\), particularly after applying refinement
procedures. As a rule of thumb, in order to credibly control for
confounders and meet the parallel trends assumptions, results for each
covariate and time period should not exceed .2 standard deviations
although smaller imbalance may be desired in some cases. The
\texttt{get\_covariate\_balance()} function takes the following key
arguments (see the function documentation for the full list):

\begin{itemize}

\item \code{...}: One or more \texttt{PanelMatch} object(s). 

\item \code{panel.data}: a \code{PanelData} object. This should be identical to the one passed to \code{PanelMatch()} and \code{PanelEstimate()} to ensure consistent results.

\item \code{covariates}: a character vector, specifying the names of the covariates for which the user is interested in calculating balance.

\item \code{include.unrefined}: logical. If set to \code{TRUE}, the function will calculate and return covariate balance measures for unrefined matched sets. This is helpful for assessing the improvement in covariate balance as a result of refining the matched sets.

\end{itemize}

Using the \texttt{get\_covariate\_balance()} function, we can compare
the balance measure across different refinement procedures such as the
Mahalanobis distance matching and propensity score weighting. Note that
the covariate balance measure should be much lower for matched sets
after refinement if the configuration used is effective.

\begin{CodeChunk}
\begin{CodeInput}
R> covbal <- get_covariate_balance(PM.maha, PM.ps.weight, 
+                                 panel.data = dem.panel,
+                                 covariates = c("tradewb", "y"),
+                                 include.unrefined = TRUE)
\end{CodeInput}
\end{CodeChunk}

The function returns a \texttt{PanelBalance} object, which contains
covariate balance results for each \texttt{PanelMatch} configuration
passed to it, which are stored in a list. Each list element is named
after the \texttt{PanelMatch} object to which it
corresponds.\footnote{In some contexts, users might see balance results that have a ``\textunderscore unrefined'' suffix appended to a \code{PanelMatch} object name. This means these are covariate balance results computed for unrefined matched sets.}
The standard \texttt{print()}, \texttt{plot()}, and \texttt{summary()}
methods are defined for the \texttt{PanelBalance} class, along with a
subset method. The \texttt{print()} method provides a way to quickly
examine the results of the refinement process. It provides the names of
\texttt{PanelMatch} objects, and the covariate balance measures for the
specified configuration(s) and variables over the lag window. One can
conduct a more careful analysis of these covariate balance results using
the \texttt{summary()} method. These results will include both refined
and unrefined balance levels for variables over the lag window if
\texttt{include.unrefined\ =\ TRUE}, as it is by default. This enables
users to assess the effectiveness of the specified matching and
refinement configuration(s). In the example below, we can see that the
Mahalanobis distance based refinement procedure seems to have improved
covariate balance measures, while the propensity score weighting
approach had limited effectiveness. Columns with the specified variable
names reflect post-refinement covariate balance values, while columns
with \texttt{\_unrefined} suffixes reflect pre-refinement balance
values. Time periods over the lag window are indexed by row, using the
format \texttt{t\_l} as a naming scheme.

\begin{CodeChunk}
\begin{CodeInput}
R> print(covbal)
\end{CodeInput}
\begin{CodeOutput}

==============================
PM.maha 
==============================

--- QOI: att ---
         tradewb          y
t_4  0.033144720 0.09725542
t_3 -0.032415036 0.10669214
t_2  0.007162189 0.07108454
t_1  0.116077872 0.04641091
t_0  0.120572693 0.03583203

==============================
PM.ps.weight 
==============================

--- QOI: att ---
       tradewb          y
t_4 0.24713734 0.42302119
t_3 0.18390406 0.26826780
t_2 0.12460414 0.18135719
t_1 0.06142984 0.03086512
t_0 0.00554273 0.02157618
\end{CodeOutput}
\begin{CodeInput}
R> summary(covbal)
\end{CodeInput}
\begin{CodeOutput}
$PM.maha
    tradewb_unrefined y_unrefined      tradewb          y
t_4       -0.10344989  0.26326816  0.033144720 0.09725542
t_3       -0.21683623  0.18343654 -0.032415036 0.10669214
t_2       -0.22188279  0.08612944  0.007162189 0.07108454
t_1       -0.09402417 -0.02126611  0.116077872 0.04641091
t_0       -0.09657564 -0.03184226  0.120572693 0.03583203

$PM.ps.weight
    tradewb_unrefined y_unrefined    tradewb          y
t_4       -0.10344989  0.26326816 0.24713734 0.42302119
t_3       -0.21683623  0.18343654 0.18390406 0.26826780
t_2       -0.22188279  0.08612944 0.12460414 0.18135719
t_1       -0.09402417 -0.02126611 0.06142984 0.03086512
t_0       -0.09657564 -0.03184226 0.00554273 0.02157618
\end{CodeOutput}
\end{CodeChunk}

One can also visualize the balance results with the \texttt{plot()}
method. Users can choose between two different types of figures,
specifying \texttt{type\ =\ "panel"}, or \texttt{type\ =\ "scatter"}.
The method will generate a figure for each configuration specified to
the \texttt{get\_covariate\_balance()}
function.\footnote{If \texttt{type = "panel"} and \texttt{include.unrefined.panel = TRUE}, then two plots will be generated for each \code{PanelMatch} configuration. First, there will be a set of plots for the refined matched sets. Then, there will be a set of plots associated with the unrefined matched sets.}
In Sections \ref{matching} and \ref{weighting} we created two different
\texttt{PanelMatch()} configurations. The two specifications are very
similar, differing only in their refinement method: one uses a matching
based approach based on Mahalanobis distance, and the other uses
propensity score weighting. Figure \ref{fig:balance_comparison} shows
the results of these refinement configurations. Each plot shows the
covariate balance for the \texttt{tradewb} and \texttt{y} variables over
the lag window, calculated by the \texttt{get\_covariate\_balance()}
function. The first plot on the left shows the calculated covariate
balance values before any refinement is applied.

Refinement using Mahalanobis distance based matching meaningfully
improved covariate balance, with calculated values comfortably below .2
standard deviations. In contrast, the propensity score weighting method
did not improve balance to acceptable levels.

\begin{CodeChunk}
\begin{CodeInput}
R> plot(covbal,
+      type = "panel", 
+      ylim = c(-.5, .6), 
+      main = "Covariate Balance")
\end{CodeInput}
\end{CodeChunk}

\begin{CodeChunk}
\begin{figure}[t]

{\centering \includegraphics{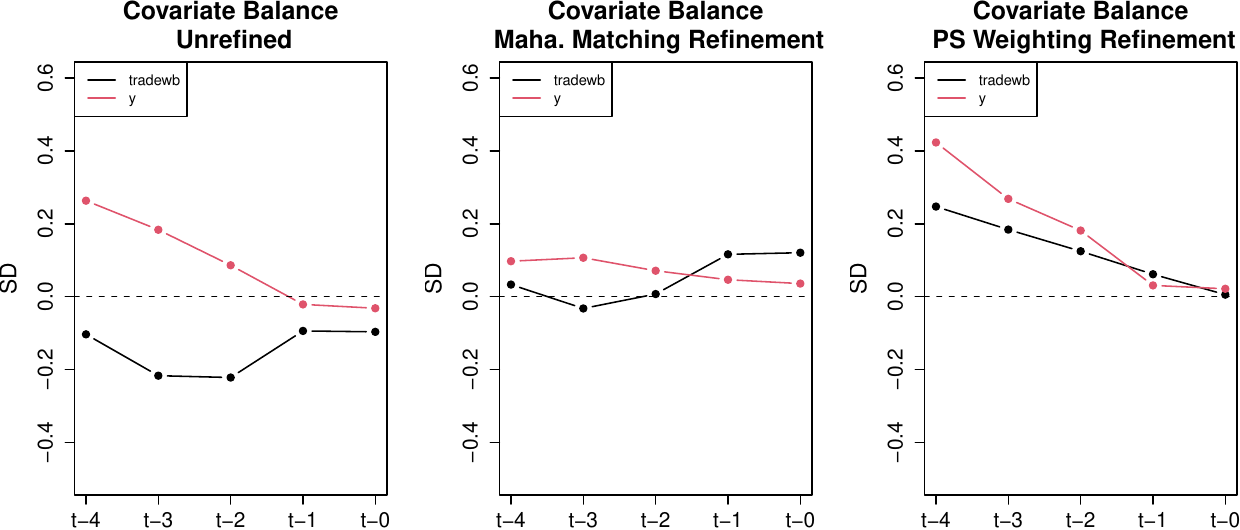} 

}

\caption{\textbf{Visualization of Covariate Balance for Unrefined and Refined Matched Sets.} Covariate balance results for unrefined matched sets are shown on the plot on the left hand side of the figure. Refining the matched sets using the Mahalanobis distance based method substantively improves covariate balance for both variables over the entire lag window, with covariate balance levels below .2 standard deviations. The propensity score weighting method did not meaningfully improve covariate balance. Users should only proceed with further analysis if covariate balance levels are acceptable.}\label{fig:balance_comparison}
\end{figure}
\end{CodeChunk}

Setting \texttt{type\ =\ "scatter"} in the \texttt{plot()} method offers
another way to assess the impact of refinement. This does require that
unrefined matched set covariate balance measures are calculated using
\texttt{get\_covariate\_balance()}. Setting \texttt{type\ =\ "scatter"}
generates a scatter plot with the following characteristics. Each point
on the plot represents a specific covariate at a particular time period
in the lag window from \(t-L\) to \(t-1\). The horizontal axis
represents the covariate balance for this particular variable and time
period before refinement is applied, while the vertical axis represents
the post-refinement balance value.

The code below shows an example of the balance scatter plot, and Figure
\ref{fig:scatter} displays the results. Here, we are visualizing the
effectiveness of the Mahalanobis distance based refinement method with
the configuration specified in Section \ref{matching}. We can observe
that the points are largely below the line \(y = x\) and below .2 on the
y-axis. This means that refinement improved covariate balance to
reasonable levels. Given these results, we proceed to calculate
estimates using \texttt{PanelEstimate()} for the \texttt{PanelMatch()}
configuration using the refinement method based on Mahalanobis distance.

\begin{CodeChunk}
\begin{CodeInput}
R> covbal <- get_covariate_balance(PM.maha, 
+                                 panel.data = dem.panel,
+                                 covariates = c("tradewb", "y"),
+                                 include.unrefined = TRUE)
R> 
R> plot(covbal, type = "scatter", ylim = c(0, .6), xlim = c(0, .6))
\end{CodeInput}
\begin{figure}[t]

{\centering \includegraphics[width=0.7\linewidth]{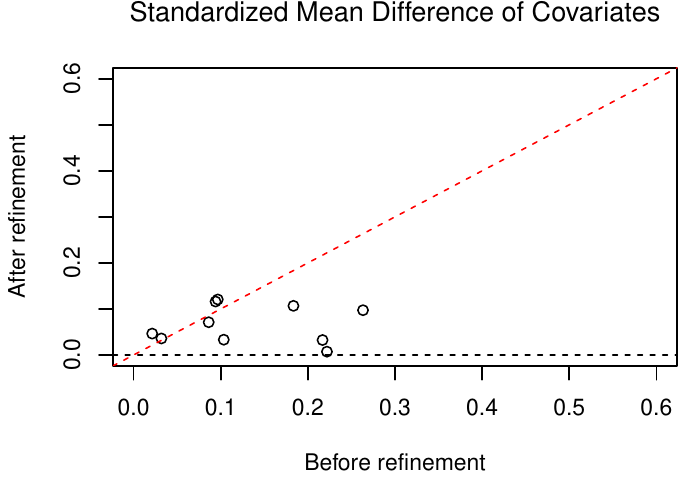} 

}

\caption{\textbf{Scatter Plot Visualizing Covariate Balance Before and After Refinement.} Each point on the plot represents the calculated measure of covariate balance defined in Equation \ref{eq:balance} for a particular variable at a particular time. The horizontal axis represents pre-refinement balance levels and the vertical axis represents post-refinement values after using the Mahalanobis distance based method. The points fall below the line $y=x$ and below .2 standard deviations on the y-axis, meaning that refinement meaningfully improved covariate balance values.}\label{fig:scatter}
\end{figure}
\end{CodeChunk}

\section{Calculating Estimates and Standard Errors with
PanelEstimate()}\label{calculating-estimates-and-standard-errors-with-panelestimate}

After evaluating the results of matching and refinement, users can then
proceed to calculate estimates and standard errors for the quantity of
interest using the \texttt{PanelEstimate()} function. The package
includes a variety of relevant diagnostic and visualization functions to
help researchers evaluate and interpret their results.

\subsection{Estimating the Quantity of
Interest}\label{estimating-the-quantity-of-interest}

The \texttt{PanelEstimate()} function calculates estimates for the
specified quantity of interest (QOI) and lead periods. Users have the
option of generating point estimates for each individual lead period
(i.e., each value of the \texttt{lead} argument in \(F\)) or one pooled
estimate for the average effect over the entire lead window. To obtain
pooled estimates, users should specify \texttt{pooled\ =\ TRUE} to
\texttt{PanelEstimate()}.

As shown in by \citet{imai2021matching}, ATT estimates can be calculated
using the following: \[
\widehat\delta_{ATT}(F,L) = \sum_{i=1}^N\sum_{t=1}^T W_{it}^\ast Y_{it}
/\sum_{i=1}^N\sum_{t=1}^T D_{it}
\] where \[
W^*_{it} = \sum_{i ^\prime = 1}^N \sum_{t^\prime = 1}^T D_{i^\prime t^\prime} v_{it}^{i^\prime t^\prime}
\] and \[v_{it}^{i^\prime t ^\prime}=
    \begin{cases}
        1 & \text{if } (i,t) = (i^\prime, t^\prime + F)\\
        -1 & \text{if } (i,t) = (i^\prime, t^\prime - 1)\\
        -w_{i^\prime t^ \prime}^{i} & \text{if } i \in \mathcal{M}_{i^\prime t^\prime}, t= t^\prime + F\\
        w_{i^\prime t^ \prime}^{i} & \text{if } \text{if } i \in \mathcal{M}_{i^\prime t^\prime}, t= t^\prime + F\\
        0 & \text{otherwise}
    \end{cases}
\]

For the ATT,
\(D_{it}=X_{it} (1-X_{i,t-1})\cdot\mathbf{1}\{|\mathcal{M}_{it}| > 0 \}\),
and \(w_{it}^{i^\prime}\) represents the non-negative normalized weights
calculated in the refinement stage (described in Sections \ref{matching}
and \ref{weighting}) such that \(w_{it}^{i^\prime} \ge 0\) and
\(\sum_{i^\prime \in \mathcal{M}_{it}} w_{it}^{i^\prime}=1\). In other
words, \(D_{it} = 1\) if and only if observation \((i,t)\) changes from
the control status at \(t-1\) to the treated status at \(t\) and has at
least one matched control unit.

The formula used to calculate the ART is similar, differing only in the
definition of \(D_{it}\). In the case of the ART, \(D_{it}\) = 1 if and
only if observation \((i,t)\) reverts from treatment status at \(t-1\)
to control status at \(t\) and has a non-empty matched set. Formulas for
the ATC and ATE can be derived analogously. See Appendix
\ref{other-estimators} for a more complete explanation.

\subsection{Standard Errors}\label{standard-errors}

Users must specify a method of calculating standard errors to the
\texttt{se.method} argument of the \texttt{PanelEstimate()} function.
The package supports a bootstrap-based method for all quantities of
interest, and two analytical methods that calculate the conditional and
unconditional standard errors for the ATT, ART, and ATC. These are
specified to the function's \texttt{se.method} argument as
\texttt{"bootstrap"}, \texttt{"conditional"}, and
\texttt{"unconditional"}, respectively.

\subsubsection{Bootstrapped Standard Errors}\label{bootstrappedSE}

The bootstrap method for obtaining standard errors works by creating
\(N\) block-bootstrap samples where each unit is a block, calculating
the QOI-specific estimate, \(\hat{\delta}(F,L)\), from each sample, and
then finding the \(\frac{\alpha}{2}\) and \(1 - \frac{\alpha}{2}\)
percentiles of the bootstrapped estimates. This method is available for
all specified quantities of interest, pooled or unpooled. Users specify
the number of bootstrap iterations to the \texttt{number.iterations}
argument.

Furthermore, the package implements a parallelized version of the
bootstrapping procedure using the \texttt{doParallel} and
\texttt{foreach} packages \citep[\citet{foreach}]{doparallel}. Users can
leverage this functionality by specifying \texttt{parallel\ =\ TRUE},
and the number of cores to be used for the calculations to the
\texttt{num.cores} argument. Note that if \texttt{se.method} is set to
anything other than \texttt{"bootstrap"}, these arguments will have no
effect.

\subsubsection{Analytical Standard
Errors}\label{analytical-standard-errors}

When the quantity of interest is the ATT, ART, or ATC, two analytical
methods for calculating standard errors are available. Conditional
standard errors are calculated assuming independence across units but
not across time. Formally, as defined in \citet{imai2021matching}:
\(A = \sum_{i=1}^N A_i\) with
\(A_i \ = \ \sum_{t=1}^T W_{it}^\ast Y_{it}\) and
\(B = \sum_{i=1}^N B_i\) with \(B_i = \sum_{t=1}^T D_{it}\). Then,
\begin{equation*}
  \mathbb{V}\left(\hat\delta(F, L) \mid \mathbf{D}\right) \ = \ \frac{N^\ast
    \mathbb{V}(A_i)}{B^2}
\end{equation*} where \(N^\ast\) is the total number of units with at
least one non-zero weight.

Alternatively, users may opt to calculate unconditional standard errors,
which make no such assumptions about independence across units. For
this, the following first-order Taylor approximation for the asymptotic
variance is used: \begin{equation}
  \mathbb{V}\left(\hat\delta(F, L) \right) \ = \  \mathbb{V}\left(\frac{A}{B}\right) \ \approx \ \frac{1}{\mathbb{E}(B)^2} \left\{\mathbb{V}(A) -
  2\frac{\mathbb{E}(A)}{\mathbb{E}(B)}\text{Cov}(A,B)  + \frac{\mathbb{E}(A)^2}{\mathbb{E}(B)^2}\mathbb{V}(B)\right\}
\end{equation}

Users will often find conditional standard errors to be smaller given
the assumed independence, relative to the other methods. Unconditional
and bootstrapped standard errors will often be comparable in size.
However, users should be careful about the credibility of the required
unit independence assumption, as well as the availability of each method
for their desired specification.

\subsection{Using PanelEstimate():
Examples}\label{using-panelestimate-examples}

\begin{table}[ht]
    \centering
    \begin{tabular}{p{0.2\linewidth} | p{0.75\linewidth}}
      Argument Name  & Description \\ \hline
      \code{sets} & A \code{PanelMatch} object \\
      \hline
      \code{panel.data} & A \code{PanelData} object with the time series cross sectional data \\
      \hline
      \code{se.method} & Method used for calculating standard errors, provided as a character string. Users must choose between \code{"bootstrap," "conditional"}, and \code{"unconditional"} methods. When the \code{qoi} provided to \code{PanelMatch()} is \code{"att", "atc"}, or \code{"art"}, all methods are available. Only \code{"bootstrap"} is available for the ATE. \\
      \hline
      \code{moderator} & The name of a categorical moderating variable, provided as a character string. This is an optional argument. If provided,the returned object will be a list of \code{PanelEstimate} objects.  \\
      \hline
      \code{pooled} &  This is an optional, logical argument specifying whether or not estimates and standard errors should be pooled across all lead periods. This is only available for \code{se.method = ``bootstrap.''}\\
      \hline
    \end{tabular}
    
    \caption{\label{tab:pe-parameters} \textbf{Main PanelEstimate() Arguments.} After finding an acceptable configuration and evaluating their results, users should provide a \code{PanelMatch} object to \code{PanelEstimate}, along with the original data set, and an appropriate method for calculating standard errors. }
\end{table}

Users must specify the following to \texttt{PanelEstimate()}: a
\texttt{PanelMatch} object, the data as a \texttt{data.frame} object,
and a method of calculating standard errors. These are provided to the
\texttt{sets}, \texttt{data}, and \texttt{se.method} arguments,
respectively, as described in Table \ref{tab:pe-parameters}. The
\texttt{PanelEstimate()} function calculates point estimates and
standard errors, according to the user's specification. Users may also
specify a moderating variable, adjust the confidence level, and change
the number of bootstrap iterations where appropriate. This section will
contain examples of how to use this function, as well as associated
methods and diagnostic functions for evaluating results.

The \texttt{PanelEstimate()} function returns a \texttt{PanelEstimate}
object. These objects are lists containing the estimates and standard
errors for each lead period, along with some associated data about the
specification. \texttt{PanelEsimate} objects have custom
\texttt{print()}, \texttt{summary()} and \texttt{plot()} methods
defined, both of which return and/or visualize data about the point
estimates and standard errors for each specified lead period.

In Section \ref{diag}, we found a \texttt{PanelMatch()} configuration
using a Mahalanobis distance based refinement procedure that generated
matched sets with acceptable sizes and levels of covariate balance. In
the following code, we calculate point estimates and standard errors for
the ATT using this specification and examine the results of
\texttt{PanelEstimate()} with the \texttt{summary()} method, which
provides some additional information.

\begin{CodeChunk}
\begin{CodeInput}
R> PE.bootstrap <- PanelEstimate(sets = PM.maha,
+                     panel.data = dem.panel,
+                     se.method = "bootstrap")
R> summary(PE.bootstrap)
\end{CodeInput}
\begin{CodeOutput}
       estimate std.error      2.5%     97.5%
t+0 -1.25055072 0.8580674 -3.026595 0.2295593
t+1 -0.78290335 1.3650242 -3.584488 1.8495049
t+2  0.05796497 1.8354613 -3.672954 3.6745426
\end{CodeOutput}
\end{CodeChunk}

Users can also specifically extract just point estimates and confidence
interval boundaries using the \texttt{estimates()} and
\texttt{confint()}
methods.\footnote{The \code{estimates()} method is intended to mirror the behavior of the familiar \code{coef()} method. Since the point estimates are not technically coefficient values in the package, a different nomenclature is employed.}

\begin{CodeChunk}
\begin{CodeInput}
R> confint(PE.bootstrap)
\end{CodeInput}
\begin{CodeOutput}
        2.5 %    97.5 %
t+0 -3.026595 0.2295593
t+1 -3.584488 1.8495049
t+2 -3.672954 3.6745426
\end{CodeOutput}
\begin{CodeInput}
R> estimates(PE.bootstrap)
\end{CodeInput}
\begin{CodeOutput}
        t+0         t+1         t+2 
-1.25055072 -0.78290335  0.05796497 
\end{CodeOutput}
\end{CodeChunk}

Users can easily visualize calculated point estimates and confidence
intervals from \texttt{PanelEstimate()} using the \texttt{plot()}
method. Users can also customize plots by specifying additional,
optional graphical parameters to be passed on to
\texttt{graphics::plot()}.

\begin{CodeChunk}
\begin{CodeInput}
R> plot(PE.bootstrap,
+       ylim = c(-4,4),
+       main = "Estimated ATT (bootstrap)")
\end{CodeInput}
\begin{figure}[t]

{\centering \includegraphics[width=0.7\linewidth]{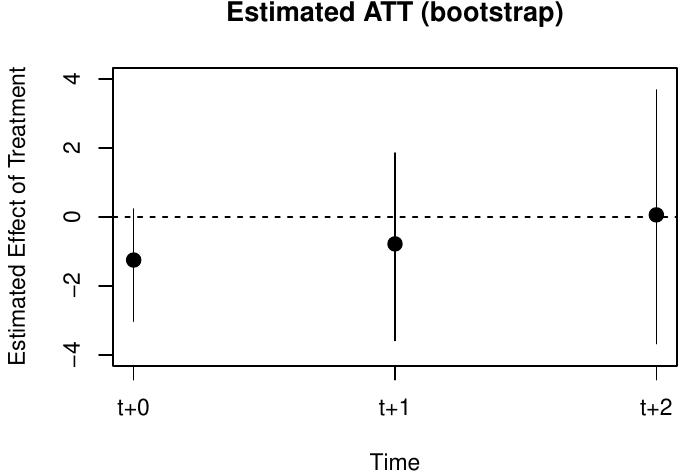} 

}

\caption{\textbf{Visualizing \code{PanelEstimate()} Results.} Users can use the \code{plot()} method for \code{PanelEstimate} objects to visualize the point estimates and confidence intervals for the specified quantity of interest over the lead window. Since \code{lead = 0:2} was specified in the corresponding \code{PanelMatch()} call with \texttt{pooled = FALSE}, results are calculated and displayed for $t+0$, $t+1$ and $t+2$, where $t$ is the time of treatment. For this specification, the estimated effects of democratization on country GDP are not statistically significant at the 95\% confidence level.}\label{fig:unnamed-chunk-16}
\end{figure}
\end{CodeChunk}

\texttt{PanelEstimate()} can handle a single, categorical moderating
variable. When a moderating variable is specified, a list of
\texttt{PanelEstimate} objects are returned. There will be a separate
\texttt{PanelEstimate} object in the list for each value of the
moderating variable, and the name of each element in the returned list
will reflect this value.

\begin{CodeChunk}
\begin{CodeInput}
R> # add simple moderating variable
R> dem$moderator <- ifelse(dem$wbcode2 > 100, "A", "B")
R> dem.panel <- PanelData(panel.data = dem, 
+                        unit.id = "wbcode2", 
+                        time.id = "year", 
+                        treatment = "dem", 
+                        outcome = "y")
R> 
R> PM.maha <- PanelMatch(panel.data = dem.panel, 
+                       lag = 4, 
+                       refinement.method = "mahalanobis",
+                       match.missing = FALSE, 
+                       covs.formula = ~ I(lag(tradewb, 0:4)) + 
+                                        I(lag(y, 1:4)),
+                       size.match = 5, 
+                       qoi = "att", 
+                       lead = 0:2,
+                       use.diagonal.variance.matrix = TRUE,
+                       forbid.treatment.reversal = FALSE)
R> 
R> PE.moderator <- PanelEstimate(sets = PM.maha, 
+                               panel.data = dem.panel,
+                               se.method = "bootstrap",
+                               moderator = "moderator")
\end{CodeInput}
\end{CodeChunk}

\subsection{Estimation Diagnostics}\label{estimation-diagnostics}

The package also includes additional diagnostic functions for evaluating
results. Users can examine matched set level estimates for the quantity
of interest using \texttt{get\_set\_treatment\_effects().} This enables
users to verify the credibility of reporting the ATT, ART, ATC, or ATE,
ensuring that the underlying distribution of observation level estimates
does not reveal any concerning heterogeneity.

The \texttt{get\_set\_treatment\_effects()} function returns a list
equal in length to the number of lead periods specified to the
\texttt{lead} argument. Each element in the list is a vector of the
matched set level effects. Each vector contains a result for every
treated \((i,t)\) with a non-empty matched set and \texttt{NA}s for
those with empty matched sets. In the following code, we use the
function to calculate matched set level treatment effects for
democratization on national GDP at times \(t\) and \(t+1\) and plot the
results for \(F = 0\) (i.e., for time \(t\)). The distribution looks
slightly left-skewed, but mostly normal in shape, verifying the
credibility of the ATT.

\begin{CodeChunk}
\begin{CodeInput}
R> ind.results <- get_set_treatment_effects(pm.obj = PM.maha, 
+                                          panel.data = dem.panel, 
+                                          lead = 0:1)
R> hist(ind.results[[1]], 
+      xlim = c(-40, 20),
+      ylim = c(0, 50),
+      main = "Distribution of Matched Set Treatment Effects\n F = 0",
+      xlab = "Treatment Effect")
\end{CodeInput}
\begin{figure}

{\centering \includegraphics[width=0.7\linewidth]{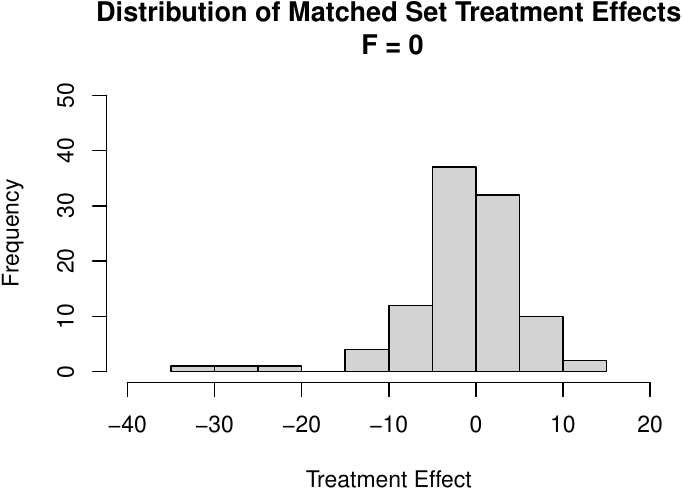} 

}

\caption{\textbf{Distribution of Matched Set Level Treatment Effects.} The \code{get\_set\_treatment\_effects()} function allows users to obtain individual, matched set level treatment effect estimates. This enables users to inspect the underlying distribution of effect estimates to validate their results for the specified quantity of interest. This figure shows a roughly normal distribution, lending credibility to the interpretation of the ATT.}\label{fig:unnamed-chunk-18}
\end{figure}
\end{CodeChunk}

Users may also conduct placebo tests with \texttt{placebo\_test()}.
Specifically, users can evaluate the effect of treatment at time \(t\)
on the difference in the outcome between the treated and control units
for the periods at \(t-2\) vs.~\(t-1\), \(t-3\) vs.~\(t-1\), and so on.
Users must specify \texttt{placebo.test\ =\ TRUE} to
\texttt{PanelMatch()} to use this
functionality.\footnote{Users may notice small differences in results returned by \code{PanelMatch()} when \code{placebo.test = TRUE} versus when \code{placebo.test = FALSE}, even if the specifications are otherwise identical. This is because placebo tests require the presence of outcome data for units over the lag window, which is not a default requirement. As a result, when \code{placebo.test = TRUE}, the size and number of matched sets might be reduced.}

Users should consider the utility and interpretation of performing a
placebo test based on their specific configuration. In particular, when
lagged outcome variables are used in the refinement process, the balance
of these lagged outcomes is expected to be good by design. Therefore, a
placebo test may be unnecessary in these situations. Alternatively, if
the refinement process excludes lagged outcome variables, it may be
beneficial for users to conduct a placebo test. This would help explore
the assumption of parallel trends, specifically expecting that the
differences in pre-treatment outcomes between treated and control groups
will remain stable over the pre-treatment lag window. Users can select a
method of standard error calculation for the test. Valid choices are the
same as for \texttt{PanelEstimate()}: \texttt{"bootstrap"},
\texttt{"unconditional"}, and \texttt{"conditional"}. When using
bootstrapped standard errors, the procedure can be parallelized by
specifying \texttt{parallel\ =\ TRUE} and the \texttt{num.cores}
argument, similar to the \texttt{PanelEstimate()} function, as described
in Section \ref{bootstrappedSE}. For convenience, users can also conduct
a placebo test by setting \texttt{include.placebo.test\ =\ TRUE} when
running \texttt{PanelEstimate()}. This will include the results of the
placebo test as an additional item in the returned
\texttt{PanelEstimate} object.

\begin{CodeChunk}
\begin{CodeInput}
R> PM.results <- PanelMatch(lag = 4, 
+                           refinement.method = "mahalanobis",
+                           panel.data = dem.panel,
+                           match.missing = TRUE,
+                           covs.formula = ~ I(lag(tradewb, 1:4)),
+                           size.match = 5, qoi = "att",
+                           lead = 0:4, 
+                           forbid.treatment.reversal = FALSE,
+                           placebo.test = TRUE)
R> placebo_test(pm.obj = PM.results, 
+              panel.data = dem.panel, 
+              number.iterations = 100, 
+              plot = TRUE, 
+              se.method = "bootstrap")
\end{CodeInput}
\begin{figure}[H]

{\centering \includegraphics{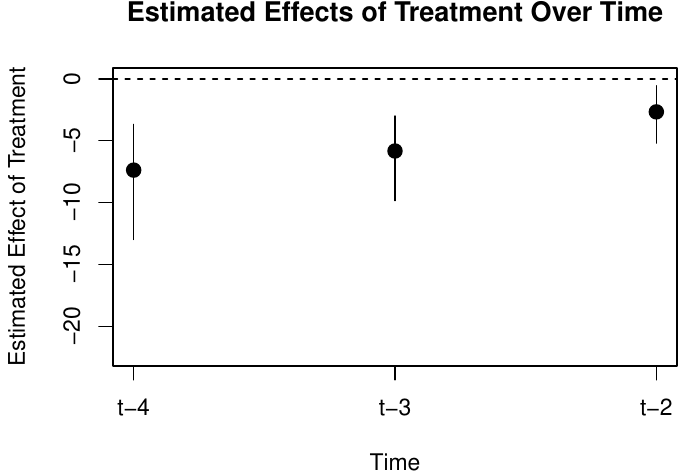} 

}

\caption{\textbf{Visualizing Results of Placebo Test.} Users can calculate and visualize the results of a placebo test using \code{placebo\_test()}. Users must indicate their desire to conduct a placebo test when specifying a \code{PanelMatch()} configuration, as this places additional restrictions on the matching and refinement procedures. The results of this trivial placebo test, showing that estimates are statistically different from zero, suggest the research design needs further refinement.}\label{fig:unnamed-chunk-19}
\end{figure}
\end{CodeChunk}

\section{Conclusion}\label{conclusion}

\textbf{PanelMatch} is an R package that implements a set of
methodological tools to help researchers apply matching methods for
causal inference on panel data with binary treatments. The package is
designed to help users follow an intuitive pipeline for visualizing and
analyzing their data and evaluating the results with a variety of
helpful and flexible functions.

Many possible extensions remain for future work. The package could be
modified to accommodate user-specified distance calculation methods for
refinement. Additional features like calipers could be added to give
further flexibility and control over matching and refinement procedures.
Furthermore, the framework of \citet{imai2021matching} could be extended
to data with continuous treatments or allow for spillover effects, as
these are active and exciting areas of research
\citep{colangelo_double_2022, dhaultfoeuille_nonparametric_2023, tubbicke_entropy_2022, vegetabile_nonparametric_2021, wu_matching_2022, callaway_difference_differences_2021}.
These developments could be integrated into future versions of the
package.

\section{Appendix}\label{appendix}

\subsection{Formal Definitions of ATC and ATE}\label{other-qoi}

The ATC is defined as: \begin{eqnarray}
\delta_{ATC}(F,L) & = &\mathbb{E}\left\{Y_{i,t+F}\left(X_{it} = 1, X_{i,t-1}=0,
  \{X_{i,t-\ell}\}_{\ell=2}^{L}\right)\right. - \nonumber \\ & & \hspace{.5in} \left. Y_{i,t+F}
  \left(X_{it} = 0, X_{i,t-1}=0, \{X_{i,t-\ell}\}_{\ell=2}^{L}\right) \mid
  X_{it} = 0, X_{i,t-1} = 0\right\}. \label{eq:qoi.atc}
\end{eqnarray} The ATE is defined as: \begin{eqnarray}
  \delta_{ATE}(F,L) 
& = & \mathbb{E}\left\{Y_{i,t+F}\left(X_{it} = 1, X_{i,t-1}=0, \{X_{i,t-\ell}\}_{\ell=2}^{L}\right)\right. - \nonumber \\
& & \hspace{.5in} \left. Y_{i,t+F} \left(X_{it} = 0, X_{i,t-1}=0, \{X_{i,t-\ell}\}_{\ell=2}^{L}\right)\right\} 
\end{eqnarray}

\subsection{Matched Sets for the ATC and ATE}\label{atc-msets}

For the ATC, matched sets are defined as: \begin{eqnarray}
\mathcal{M}_{it}^{\textrm{ATC}} = \{i^\prime : i^\prime \ne i, X_{i^\prime t} =1,
X_{i^\prime t^\prime} = X_{i t^\prime} \textrm{ for all } t^\prime =
t-1, \dots,t-L\}
\end{eqnarray} where treated observations have \(X_{it}=0\) and
\(X_{i,t-1}=0\). When the ATE is specified, matched sets for the ATT and
ATC are identified separately.

\subsection{Customizing matched.set objects}\label{custom-msets}

The package allows for users to specify custom \texttt{matched.set}
objects, although this should be done with care. For instance,
researchers might have a specific, substantively motivated matching or
refinement scheme that is not implemented by the package. As long as
users' custom specifications follow the described structure and do not
violate any methodological expectations, the package will accommodate
such requests.

The following code demonstrates a simple example, creating unrefined
matched sets, and specifying custom weights for control units matched to
unit id = 133, which received treatment in 2010.

\begin{CodeChunk}
\begin{CodeInput}
R> PM.results <- PanelMatch(lag = 4, 
+                          qoi = "att", 
+                          lead = 0,
+                          refinement.method = "none",
+                          panel.data = dem.panel)
R> mset <- PM.results[["att"]][["133.2010"]]
R> print(mset)
\end{CodeInput}
\begin{CodeOutput}
[1] 116 192
attr(,"treatment.change")
[1] 1
attr(,"control.change")
[1] 0 0
attr(,"weights")
116 192 
0.5 0.5 
\end{CodeOutput}
\begin{CodeInput}
R> attr(mset, "weights") <- setNames(c(.1, .9), c(116, 192))
R> print(mset)
\end{CodeInput}
\begin{CodeOutput}
[1] 116 192
attr(,"treatment.change")
[1] 1
attr(,"control.change")
[1] 0 0
attr(,"weights")
116 192 
0.1 0.9 
\end{CodeOutput}
\end{CodeChunk}

\subsection{Formal Definitions of ATC, and ATE
Estimators}\label{other-estimators}

\subsubsection{ATC}\label{atc}

The estimator for the ATC is defined as \begin{align*}
\widehat\delta_{ATC}(F,L) = &\frac{1}{\sum_{i=1}^N \sum_{t=L+1}^{T-F} D_{it}} \sum_{i=1}^N \sum_{t=L+1}^{T-F}  D_{it}
\left( \sum_{i^\prime \in \mathcal{M}_{it}} w_{it}^{i^\prime} (Y_{i^\prime,t+F} - Y_{i^\prime,t-1}) - (Y_{i,t+F - Y_{i,t-1}}) \right)
\end{align*} but can be expressed as
\(-\sum_{i=1}^N\sum_{t=1}^T W_{it}^\ast Y_{it} /\sum_{i=1}^N\sum_{t=1}^T D_{it}\),
just as the ATT and ART, as described in \citet{imai2021matching}. For
the ATC, \(D_{it} = 1\) when \(X_{i,t} = X_{i,t-1} = 0\) and the
corresponding matched set for observation (i,t) is non-empty.

\subsubsection{ATE}\label{ate}

Define \(D^{att}_{it} = 1\) when \(X_{it-1} = 0, X_{it} = 1\) and there
is at least 1 matched control unit for observation \((i,t)\). Similarly,
define \(D^{atc}_{it} = 1\) when \(X_{it-1} = 0, X_{it} = 0\) and the
corresponding matched set for observation \((i,t)\) is non-empty. Then,
let \(\boldsymbol{D_{att}} = \sum_{i=1}^N\sum_{t=1}^T D^{att}_{it}\),
\(\boldsymbol{D_{atc}} = \sum_{i=1}^N\sum_{t=1}^T D^{atc}_{it}\), and
\(\boldsymbol{D_{ate}} = \boldsymbol{D_{att}} + \boldsymbol{D_{atc}}\).
Then, \[
\widehat\delta_{ate}(F, L) = \frac{\widehat\delta_{att}(F, L) \sum_{i=1}^N\sum_{t=1}^T D^{att}_{it} + \widehat\delta_{atc}(F, L) \sum_{i=1}^N\sum_{t=1}^T D^{atc}_{it}}{\sum_{i=1}^N\sum_{t=1}^T D^{att}_{it} + \sum_{i=1}^N\sum_{t=1}^T D^{atc}_{it}} 
\] and \[
\widehat\delta_{ate}(F, L) = \frac{\widehat\delta_{att}(F, L) \boldsymbol{D_{att}} + \widehat\delta_{atc}(F, L)\boldsymbol{D_{atc}} }{\boldsymbol{D_{att}} + \boldsymbol{D_{atc}}} 
\]

\subsection{Further details about summarizing and printing matched.set
objects}\label{mset-other-methods}

The \texttt{summary()} method will provide users with information about
the sizes of matched sets. The function accepts one non-standard logical
argument, \texttt{verbose} (default is \code{FALSE}), which controls how
much information is calculated and returned to the user. It returns a
list with the following elements:

\begin{itemize}
\item \code{overview}: A \code{data.frame} object containing information about the treated units (unit identifier, time of treatment), and the number of matched control units with weights zero and above. This is returned for both \code{verbose = TRUE} and \code{verbose = FALSE}
\item \code{set.size.summary}: a \code{summary} object summarizing the minimum, maximum, and IQR of matched set sizes. It is only returned for \code{verbose = TRUE}.
\item \code{number.of.treated.units}: The number of unit, time pairs that are considered to be ``treated'' observations. It is only returned for \code{verbose = TRUE}.
\item \code{num.units.empty.set}: The number of treated observations that were not able to be matched to any control units. It is only returned for \code{verbose = TRUE}.
\item \code{lag}: The size of the lag window used for matching on treatment history. It is only returned for \code{verbose = TRUE}.
\end{itemize}

The \texttt{print()} method displays information about the sizes of
matched sets. By default, it provides details about the treated
observations (specifically, the time of treatment and the ID of the
treated observation), as well as the sizes of matched sets. The method
accepts a few important parameters. By default, only information about
the first 5 matched sets is shown. Users can increase or decrease the
number of previewed matched sets by adjusting \texttt{n}. Users can also
show information about all matched sets with \texttt{show.all\ =\ TRUE},
or get even more verbose output showing the unit identifiers of all
controls in matched sets with \texttt{verbose\ =\ TRUE}. Both options
are \texttt{FALSE} by default.

\begin{CodeChunk}
\begin{CodeInput}
R> pm.results <- PanelMatch(panel.data = dem.panel,
+                          lag = 4,
+                          refinement.method = "none",
+                          match.missing = FALSE,
+                          qoi = "att",
+                          lead = 0:2)
R> 
R> msets <- extract(pm.results)
R> summary(msets)
\end{CodeInput}
\begin{CodeOutput}
$overview
    wbcode2 year matched.set.size
1         4 1992               74
2         4 1997                2
3         6 1973               63
4         6 1983               73
5         7 1991               81
6         7 1998                1
7        12 1992               74
8        13 2003               58
9        15 1991               81
10       16 1977               63
11       17 1991               81
12       18 1991               81
13       22 1991               81
14       25 1982               72
15       26 1985               76
16       29 2008               48
17       31 1993               65
18       34 1990               79
19       36 2000               57
20       38 1992               74
21       40 1990               79
22       40 1996                1
23       40 2002                1
24       41 1991               81
25       45 1993               65
26       47 1999               59
27       50 1978               63
28       52 1979               60
29       55 1978               63
30       56 1992               74
31       57 1995               57
32       59 1990                1
33       64 1995               57
34       65 1970               58
35       65 1979               60
36       65 1996               57
37       70 1994               58
38       70 2005                1
39       72 1975               61
40       73 1984               74
41       75 1966               46
42       75 1986               77
43       78 1992               74
44       80 1982               72
45       81 2000               57
46       82 2006               55
47       83 1990               79
48       84 1999               59
49       96 2002               58
50       97 2005               56
51      101 1988               80
52      104 2005               56
53      105 2004               58
54      109 1993               65
55      110 1993               65
56      112 1993               65
57      115 1994               58
58      116 1993               65
59      118 1997               58
60      119 1991               81
61      120 1992               74
62      123 1993               65
63      124 1994               58
64      125 2007               54
65      128 1994               58
66      133 1991               81
67      134 1979               60
68      134 1999               59
69      135 1990               79
70      138 1991               81
71      138 2006               55
72      141 1972               63
73      141 1988               80
74      141 2008               48
75      142 1994               58
76      143 1980               61
77      144 1987               78
78      149 1976               63
79      150 1993               65
80      154 1990               79
81      155 1993               65
82      158 1965               40
83      158 1986               77
84      159 2000               57
85      162 2004               58
86      163 1996               57
87      163 2001               59
88      164 1982               72
89      167 1988               80
90      168 1993               65
91      169 1992               74
92      177 1974               62
93      177 1978                1
94      177 2008                1
95      183 1983                2
96      187 1994               58
97      188 1985               76
98      199 1994               58
99      201 1991               81
100     202 1978               63
101      17 2009                0
102      70 1999                0
103      82 1994                0
104     109 1999                0
105     133 1999                0
106     143 1993                0
107     167 1991                0
108     177 1992                0
109     183 1973                0

$set.size.summary
   Min. 1st Qu.  Median    Mean 3rd Qu.    Max. 
   0.00   57.00   61.00   55.26   74.00   81.00 

$number.of.treated.units
[1] 109

$num.units.empty.set
[1] 9

$lag
[1] 4
\end{CodeOutput}
\begin{CodeInput}
R> print(msets)
\end{CodeInput}
\begin{CodeOutput}
  wbcode2 year matched.set.size
1       4 1992               74
2       4 1997                2
3       6 1973               63
4       6 1983               73
5       7 1991               81
... [104 more matched set(s) not printed]
\end{CodeOutput}
\begin{CodeInput}
R> print(msets, show.all = TRUE)
\end{CodeInput}
\begin{CodeOutput}
    wbcode2 year matched.set.size
1         4 1992               74
2         4 1997                2
3         6 1973               63
4         6 1983               73
5         7 1991               81
6         7 1998                1
7        12 1992               74
8        13 2003               58
9        15 1991               81
10       16 1977               63
11       17 1991               81
12       18 1991               81
13       22 1991               81
14       25 1982               72
15       26 1985               76
16       29 2008               48
17       31 1993               65
18       34 1990               79
19       36 2000               57
20       38 1992               74
21       40 1990               79
22       40 1996                1
23       40 2002                1
24       41 1991               81
25       45 1993               65
26       47 1999               59
27       50 1978               63
28       52 1979               60
29       55 1978               63
30       56 1992               74
31       57 1995               57
32       59 1990                1
33       64 1995               57
34       65 1970               58
35       65 1979               60
36       65 1996               57
37       70 1994               58
38       70 2005                1
39       72 1975               61
40       73 1984               74
41       75 1966               46
42       75 1986               77
43       78 1992               74
44       80 1982               72
45       81 2000               57
46       82 2006               55
47       83 1990               79
48       84 1999               59
49       96 2002               58
50       97 2005               56
51      101 1988               80
52      104 2005               56
53      105 2004               58
54      109 1993               65
55      110 1993               65
56      112 1993               65
57      115 1994               58
58      116 1993               65
59      118 1997               58
60      119 1991               81
61      120 1992               74
62      123 1993               65
63      124 1994               58
64      125 2007               54
65      128 1994               58
66      133 1991               81
67      134 1979               60
68      134 1999               59
69      135 1990               79
70      138 1991               81
71      138 2006               55
72      141 1972               63
73      141 1988               80
74      141 2008               48
75      142 1994               58
76      143 1980               61
77      144 1987               78
78      149 1976               63
79      150 1993               65
80      154 1990               79
81      155 1993               65
82      158 1965               40
83      158 1986               77
84      159 2000               57
85      162 2004               58
86      163 1996               57
87      163 2001               59
88      164 1982               72
89      167 1988               80
90      168 1993               65
91      169 1992               74
92      177 1974               62
93      177 1978                1
94      177 2008                1
95      183 1983                2
96      187 1994               58
97      188 1985               76
98      199 1994               58
99      201 1991               81
100     202 1978               63
101      17 2009                0
102      70 1999                0
103      82 1994                0
104     109 1999                0
105     133 1999                0
106     143 1993                0
107     167 1991                0
108     177 1992                0
109     183 1973                0
\end{CodeOutput}
\end{CodeChunk}

\bibliography{references.bib}

\begin{thebibliography}{42}
\newcommand{\enquote}[1]{``#1''}
\providecommand{\natexlab}[1]{#1}
\providecommand{\url}[1]{\texttt{#1}}
\providecommand{\urlprefix}{URL }
\expandafter\ifx\csname urlstyle\endcsname\relax
  \providecommand{\doi}[1]{doi:\discretionary{}{}{}#1}\else
  \providecommand{\doi}{doi:\discretionary{}{}{}\begingroup
  \urlstyle{rm}\Url}\fi
\providecommand{\eprint}[2][]{\url{#2}}

\bibitem[{Abadie \emph{et~al.}(2010)Abadie, Diamond, and
  Hainmueller}]{abadie2010synthetic}
Abadie A, Diamond A, Hainmueller J (2010).
\newblock \enquote{Synthetic control methods for comparative case studies:
  Estimating the effect of California’s tobacco control program.}
\newblock \emph{Journal of the American statistical Association},
  \textbf{105}(490), 493--505.

\bibitem[{Acemoglu \emph{et~al.}(2019)Acemoglu, Naidu, Restrepo, and
  Robinson}]{acemoglu_democracy_2017}
Acemoglu D, Naidu S, Restrepo P, Robinson JA (2019).
\newblock \enquote{Democracy Does Cause Growth.}
\newblock \emph{Journal of Political Economy}, \textbf{127}(1), 47--100.

\bibitem[{Arkhangelsky \emph{et~al.}(2021)Arkhangelsky, Athey, Hirshberg,
  Imbens, and Wager}]{arkhangelsky2021syntheticdifferencedifferences}
Arkhangelsky D, Athey S, Hirshberg DA, Imbens GW, Wager S (2021).
\newblock \enquote{Synthetic Difference in Differences.}
\newblock \eprint{1812.09970},
  \urlprefix\url{https://arxiv.org/abs/1812.09970}.

\bibitem[{Ben-Michael \emph{et~al.}(2019)Ben-Michael, Feller, Rothstein
  \emph{et~al.}}]{ben2019synthetic}
Ben-Michael E, Feller A, Rothstein J, \emph{et~al.} (2019).
\newblock \enquote{Synthetic controls and weighted event studies with staggered
  adoption.}
\newblock \emph{arXiv preprint arXiv:1912.03290}, \textbf{2}.

\bibitem[{Berg\'e(2018)}]{fixest}
Berg\'e L (2018).
\newblock \enquote{Efficient estimation of maximum likelihood models with
  multiple fixed-effects: the {R} package {FENmlm}.}
\newblock \emph{CREA Discussion Papers}, (13).

\bibitem[{Borusyak \emph{et~al.}(2024)Borusyak, Jaravel, and
  Spiess}]{borusyak2024revisiting}
Borusyak K, Jaravel X, Spiess J (2024).
\newblock \enquote{Revisiting event-study designs: robust and efficient
  estimation.}
\newblock \emph{Review of Economic Studies}, p. rdae007.

\bibitem[{Branas \emph{et~al.}(2011)Branas, Cheney, MacDonald, Tam, Jackson,
  and Ten~Have}]{branas2011difference}
Branas CC, Cheney RA, MacDonald JM, Tam VW, Jackson TD, Ten~Have TR (2011).
\newblock \enquote{A difference-in-differences analysis of health, safety, and
  greening vacant urban space.}
\newblock \emph{American journal of epidemiology}, \textbf{174}(11),
  1296--1306.

\bibitem[{Butts(2021)}]{didimputation}
Butts K (2021).
\newblock \emph{didimputation: Difference-in-Differences estimator from
  Borusyak, Jaravel, and Spiess (2021)}.
\newblock \urlprefix\url{https://github.com/kylebutts/didimputation}.

\bibitem[{Butts and Gardner(2022)}]{RJ-2022-048}
Butts K, Gardner J (2022).
\newblock \enquote{did2s: Two-Stage Difference-in-Differences.}
\newblock \emph{The R Journal}, \textbf{14}, 162--173.
\newblock ISSN 2073-4859.
\newblock \doi{10.32614/RJ-2022-048}.
\newblock Https://doi.org/10.32614/RJ-2022-048.

\bibitem[{Callaway \emph{et~al.}(2021)Callaway, Goodman-Bacon, and
  Sant'Anna}]{callaway_difference_differences_2021}
Callaway B, Goodman-Bacon A, Sant'Anna PHC (2021).
\newblock \enquote{Difference-in-{Differences} with a {Continuous}
  {Treatment}.}
\newblock ArXiv:2107.02637 [econ],
  \urlprefix\url{http://arxiv.org/abs/2107.02637}.

\bibitem[{Callaway and Sant'Anna(2021{\natexlab{a}})}]{did-package}
Callaway B, Sant'Anna PH (2021{\natexlab{a}}).
\newblock \enquote{did: Difference in Differences.}
\newblock R package version 2.1.2,
  \urlprefix\url{https://bcallaway11.github.io/did/}.

\bibitem[{Callaway and Sant'Anna(2021{\natexlab{b}})}]{did-method}
Callaway B, Sant'Anna PH (2021{\natexlab{b}}).
\newblock \enquote{Difference-in-differences with multiple time periods.}
\newblock \emph{Journal of Econometrics}.
\newblock \urlprefix\url{https://doi.org/10.1016/j.jeconom.2020.12.001}.

\bibitem[{Card and Krueger(1994)}]{CardKrueger1994}
Card D, Krueger AB (1994).
\newblock \enquote{Minimum Wages and Employment: A Case Study of the Fast-Food
  Industry in New Jersey and Pennsylvania.}
\newblock \emph{American Economic Review}, \textbf{84}(4), 772--793.

\bibitem[{Colangelo and Lee(2022)}]{colangelo_double_2022}
Colangelo K, Lee YY (2022).
\newblock \enquote{Double {Debiased} {Machine} {Learning} {Nonparametric}
  {Inference} with {Continuous} {Treatments}.}
\newblock ArXiv:2004.03036 [econ],
  \urlprefix\url{http://arxiv.org/abs/2004.03036}.

\bibitem[{Croissant and Millo(2008)}]{croissant2008panel}
Croissant Y, Millo G (2008).
\newblock \enquote{Panel data econometrics in R: The plm package.}
\newblock \emph{Journal of statistical software}, \textbf{27}(2), 1--43.

\bibitem[{Doudchenko and Imbens(2016)}]{doudchenko2016balancing}
Doudchenko N, Imbens GW (2016).
\newblock \enquote{Balancing, regression, difference-in-differences and
  synthetic control methods: A synthesis.}
\newblock \emph{Technical report}, National Bureau of Economic Research.

\bibitem[{D’Haultfœuille \emph{et~al.}(2023)D’Haultfœuille, Hoderlein,
  and Sasaki}]{dhaultfoeuille_nonparametric_2023}
D’Haultfœuille X, Hoderlein S, Sasaki Y (2023).
\newblock \enquote{Nonparametric difference-in-differences in repeated
  cross-sections with continuous treatments.}
\newblock \emph{Journal of Econometrics}, \textbf{234}(2), 664--690.
\newblock ISSN 03044076.
\newblock \doi{10.1016/j.jeconom.2022.07.003}.
\newblock
  \urlprefix\url{https://linkinghub.elsevier.com/retrieve/pii/S0304407622001452}.

\bibitem[{Gardner(2022)}]{gardner2022two}
Gardner J (2022).
\newblock \enquote{Two-stage differences in differences.}
\newblock \emph{arXiv preprint arXiv:2207.05943}.

\bibitem[{Gaure(2013)}]{RJ-2013-031}
Gaure S (2013).
\newblock \enquote{lfe: Linear Group Fixed Effects.}
\newblock \emph{The R Journal}, \textbf{5}, 104--116.
\newblock ISSN 2073-4859.
\newblock Https://rjournal.github.io/.

\bibitem[{Hazlett and Xu(2018)}]{hazlett2018trajectory}
Hazlett C, Xu Y (2018).
\newblock \enquote{Trajectory balancing: A general reweighting approach to
  causal inference with time-series cross-sectional data.}
\newblock \emph{Available at SSRN 3214231}.

\bibitem[{Hirano \emph{et~al.}(2003)Hirano, Imbens, and Ridder}]{hirano}
Hirano K, Imbens G, Ridder G (2003).
\newblock \enquote{Efficient Estimation of Average Treatment Effects Using the
  Estimated Propensity Score.}
\newblock \emph{Econometrica}, \textbf{71}(4), 1307--1338.

\bibitem[{Ho \emph{et~al.}(2007)Ho, Imai, King, and Stuart}]{ho2007matching}
Ho DE, Imai K, King G, Stuart EA (2007).
\newblock \enquote{Matching as nonparametric preprocessing for reducing model
  dependence in parametric causal inference.}
\newblock \emph{Political analysis}, \textbf{15}(3), 199--236.

\bibitem[{Imai and Kim(2019)}]{imaikim19}
Imai K, Kim IS (2019).
\newblock \enquote{When Should We Use Linear Unit Fixed Effects Regression
  Models for Causal Inference with Longitudinal Data?}
\newblock \emph{American Journal of Political Science}, \textbf{63}(2),
  467--490.

\bibitem[{Imai and Kim(2021)}]{imai2021use}
Imai K, Kim IS (2021).
\newblock \enquote{On the use of two-way fixed effects regression models for
  causal inference with panel data.}
\newblock \emph{Political Analysis}, \textbf{29}(3), 405--415.

\bibitem[{Imai \emph{et~al.}(2023)Imai, Kim, and Wang}]{imai2021matching}
Imai K, Kim IS, Wang EH (2023).
\newblock \enquote{Matching Methods for Causal Inference with Time-Series
  Cross-Sectional Data.}
\newblock \emph{American Journal of Political Science}, \textbf{67}(3),
  587--605.

\bibitem[{Imai and Ratkovic(2014)}]{imairatk14}
Imai K, Ratkovic M (2014).
\newblock \enquote{Covariate Balancing Propensity Score.}
\newblock \emph{Journal of the Royal Statistical Society, {Series B}
  (Statistical Methodology)}, \textbf{76}(1), 243--263.

\bibitem[{{Jens Hainmueller} \emph{et~al.}(2011){Jens Hainmueller}, {Alexis
  Diamond}, and {Alberto Abadie}}]{Synth}
{Jens Hainmueller}, {Alexis Diamond}, {Alberto Abadie} (2011).
\newblock \enquote{Synth: An R Package for Synthetic Control Methods in
  Comparative Case Studies.}
\newblock \emph{Journal of Statistical Software}, \textbf{42}(13), 1--17.
\newblock \urlprefix\url{https://www.jstatsoft.org/v42/i13/}.

\bibitem[{Long(2020)}]{panelr}
Long JA (2020).
\newblock \emph{panelr: Regression Models and Utilities for Repeated Measures
  and Panel Data}.
\newblock R package version 0.7.3,
  \urlprefix\url{https://cran.r-project.org/package=panelr}.

\bibitem[{{Microsoft} and Weston(2022{\natexlab{a}})}]{doparallel}
{Microsoft}, Weston S (2022{\natexlab{a}}).
\newblock \emph{doParallel: Foreach Parallel Adaptor for the 'parallel'
  Package}.
\newblock R package version 1.0.17,
  \urlprefix\url{https://CRAN.R-project.org/package=doParallel}.

\bibitem[{{Microsoft} and Weston(2022{\natexlab{b}})}]{foreach}
{Microsoft}, Weston S (2022{\natexlab{b}}).
\newblock \emph{foreach: Provides Foreach Looping Construct}.
\newblock R package version 1.5.2,
  \urlprefix\url{https://CRAN.R-project.org/package=foreach}.

\bibitem[{Mou \emph{et~al.}(2022)Mou, Liu, and Xu}]{mou2022panelview}
Mou H, Liu L, Xu Y (2022).
\newblock \enquote{panelView: Panel Data Visualization in R and Stata.}
\newblock \emph{Available at SSRN 4202154}.

\bibitem[{{R Core Team}(2024)}]{stats-package}
{R Core Team} (2024).
\newblock \emph{R: A Language and Environment for Statistical Computing}.
\newblock R Foundation for Statistical Computing, Vienna, Austria.
\newblock \urlprefix\url{https://www.R-project.org/}.

\bibitem[{Roth and Sant'Anna(2021)}]{RothSantAnna2021}
Roth J, Sant'Anna PHC (2021).
\newblock \enquote{Staggered: Efficient Estimation Under Staggered Treatment
  Timing.}
\newblock \urlprefix\url{https://github.com/jonathandroth/staggered}.

\bibitem[{Roth and Sant’Anna(2023)}]{roth2023efficient}
Roth J, Sant’Anna PH (2023).
\newblock \enquote{Efficient estimation for staggered rollout designs.}
\newblock \emph{Journal of Political Economy Microeconomics}, \textbf{1}(4),
  669--709.

\bibitem[{Sant'Anna and Zhao(2020)}]{drdid}
Sant'Anna PHC, Zhao J (2020).
\newblock \enquote{Doubly Robust Difference-in-Differences Estimators.}
\newblock \emph{Journal of Econometrics}, \textbf{219}, 101--122.
\newblock \urlprefix\url{https://doi.org/10.1016/j.jeconom.2020.06.003}.

\bibitem[{Stuart(2010)}]{stuart2010matching}
Stuart EA (2010).
\newblock \enquote{Matching methods for causal inference: A review and a look
  forward.}
\newblock \emph{Statistical science: a review journal of the Institute of
  Mathematical Statistics}, \textbf{25}(1), 1.

\bibitem[{Tübbicke(2022)}]{tubbicke_entropy_2022}
Tübbicke S (2022).
\newblock \enquote{Entropy {Balancing} for {Continuous} {Treatments}.}
\newblock \emph{Journal of Econometric Methods}, \textbf{11}(1), 71--89.
\newblock ISSN 2156-6674.
\newblock \doi{10.1515/jem-2021-0002}.
\newblock
  \urlprefix\url{https://www.degruyter.com/document/doi/10.1515/jem-2021-0002/html}.

\bibitem[{Vegetabile \emph{et~al.}(2021)Vegetabile, Griffin, Coffman, Cefalu,
  Robbins, and McCaffrey}]{vegetabile_nonparametric_2021}
Vegetabile BG, Griffin BA, Coffman DL, Cefalu M, Robbins MW, McCaffrey DF
  (2021).
\newblock \enquote{Nonparametric estimation of population average dose-response
  curves using entropy balancing weights for continuous exposures.}
\newblock \emph{Health Services and Outcomes Research Methodology},
  \textbf{21}(1), 69--110.
\newblock ISSN 1387-3741, 1572-9400.
\newblock \doi{10.1007/s10742-020-00236-2}.
\newblock \urlprefix\url{http://link.springer.com/10.1007/s10742-020-00236-2}.

\bibitem[{Wickham(2016)}]{ggplot_2}
Wickham H (2016).
\newblock \emph{ggplot2: Elegant Graphics for Data Analysis}.
\newblock Springer-Verlag New York.
\newblock ISBN 978-3-319-24277-4.
\newblock \urlprefix\url{https://ggplot2.tidyverse.org}.

\bibitem[{Wu \emph{et~al.}(2022)Wu, Mealli, Kioumourtzoglou, Dominici, and
  Braun}]{wu_matching_2022}
Wu X, Mealli F, Kioumourtzoglou MA, Dominici F, Braun D (2022).
\newblock \enquote{Matching on {Generalized} {Propensity} {Scores} with
  {Continuous} {Exposures}.}
\newblock \emph{Journal of the American Statistical Association}, pp. 1--29.
\newblock ISSN 0162-1459, 1537-274X.
\newblock \doi{10.1080/01621459.2022.2144737}.
\newblock
  \urlprefix\url{https://www.tandfonline.com/doi/full/10.1080/01621459.2022.2144737}.

\bibitem[{Xu(2017)}]{xu2017generalized}
Xu Y (2017).
\newblock \enquote{Generalized synthetic control method: Causal inference with
  interactive fixed effects models.}
\newblock \emph{Political Analysis}, \textbf{25}(1), 57--76.

\bibitem[{Xu and Liu(2022)}]{gsynth}
Xu Y, Liu L (2022).
\newblock \emph{gsynth: Generalized Synthetic Control Method}.
\newblock R package version 1.2.1,
  \urlprefix\url{https://yiqingxu.org/packages/gsynth/}.

\end{thebibliography}

\end{document}